\documentclass[12pt]{article}
\usepackage{multirow}  
\usepackage{amsmath}
\usepackage{amssymb, latexsym}
 \usepackage{amsthm}
 \usepackage{caption}
\usepackage{setspace} 
\usepackage{amsthm,color}
\def\theequation{\thesection.\arabic{equation}}
\def\no{\noindent}

\def\T{{ \mathrm{\scriptscriptstyle T} }}

\def\no{\noindent}
\def\PELz{\mbox{\tiny PEL0}}
\def\PELo{\mbox{\tiny PEL1}}
\def\PEL{\mbox{\tiny PEL}}
\def\PPEL{\mbox{\tiny PPEL}}
\def\POEL{\mbox{\tiny POEL}}
\def\SELz{\mbox{\tiny SEL0}}
\def\SEL{\mbox{\tiny SEL}}
\def\PSEL{\mbox{\tiny PSEL}}

\def\BT{\mbox{\tiny BT}}
\def\U{\mbox{\tiny U}}
\def\L{\mbox{\tiny L}}
\def\N{\mbox{\tiny N}}
\def\Nj{\mbox{\tiny Nj}}
\def\Np{\mbox{\tiny Np}}
\def\Nzero{\mbox{\tiny N0}}
\def\None{\mbox{\tiny N1}}
\def\Ntwo{\mbox{\tiny N2}}
\def\Nthree{\mbox{\tiny N3}}
\def\Nft{\mbox{\tiny N14}}
\def\Ntw{\mbox{\tiny N12}}
\def\R{\scriptscriptstyle R}
\def\HT{\mbox{\tiny HT}}

\def\bfp{\mbox{\boldmath{$p$}}}

\def\mU{\mathcal{U}}
\def\mS{\mathcal{S}}
\def\mF{\mathcal{F}}

\numberwithin{equation}{section}
\theoremstyle{plain}

\begin{document}

\centerline{\large {\bf Empirical Likelihood Inference With Public-Use Survey Data}}

\bigskip

\centerline{Puying Zhao, \ J.N.K. Rao \ and \ Changbao Wu\footnote{Puying Zhao is Associate Professor, Yunnan University, China; J.N.K. Rao is Distinguished Research Professor, Carleton University, Canada;  and Changbao Wu is Professor, University of Waterloo, Canada (E-mails: {\tt pyzhao@live.cn}, \ {\tt jrao34@rogers.com} \ and \ {\tt cbwu@uwaterloo.ca}). This research was supported by grants from the Natural Sciences and Engineering Research Council of Canada (NSERC) and the Canadian Statistical Sciences Institute (CANSSI).}}

\bigskip

\bigskip

\hrule

{\small
\begin{quotation}
\no
Public-use survey data are an important source of information for researchers in social science and health studies to build statistical models and make inferences on the target finite population. This paper presents two general inferential tools through the pseudo empirical likelihood and the sample empirical likelihood methods. Theoretical results on point estimation and linear or nonlinear hypothesis tests involving parameters defined through estimating equations are established, and practical issues with the implementation of the proposed methods are discussed. Results from simulation studies and an application to the 2016 General Social Survey dataset of Statistics Canada show that the proposed methods work well under different scenarios. The inferential procedures and theoretical results presented in the paper make the empirical likelihood a practically useful tool for users of complex survey data.

\vspace{0.3cm}

\no
\end{quotation}
}

\hrule

\bigskip

\bigskip

\setcounter{section}{1}
\setcounter{equation}{0}
\no {\bf 1. Introduction}

\medskip

\no
Owen (1988) proposed the empirical likelihood approach for making inference from independent and identically distributed random samples. He showed that the empirical likelihood ratio statistic for the population mean has a standard limiting chi-squared distribution, and used this result to obtain confidence intervals for the population mean similar to the classic parametric method. Qin and Lawless (1994) demonstrated that empirical likelihood can be combined with estimating equations for statistical inferences with more general parameters. The development of empirical likelihood as a general inferential tool has been one of the major advances in statistics in the past three decades.

Empirical likelihood was in fact first introduced in the sample survey context by Hartley and Rao (1968) as the scale-load likelihood, but their focus was on point estimation of a finite population mean under simple random sampling and stratified simple random sampling. Chen and Qin (1993) studied empirical likelihood under simple random sampling using the formulation of Owen (1988), and Zhong and Rao (2000) studied empirical likelihood confidence intervals on the finite population mean under stratified simple random sampling. For general sampling designs involving unequal probability sampling with or without stratification, there have been several proposed approaches on empirical likelihood for complex surveys, including the pseudo empirical likelihood method of Chen and Sitter (1999) and Wu and Rao (2006), the population empirical likelihood method of Chen and Kim (2014), and the empirical likelihood method of Berger and Torres (2016) and Oguz-Alper and Berger (2016). However, all existing methods require the first order inclusion probabilities from the initial survey design and are developed under the setting that detailed design information is available.  In addition, the use of calibration constraints for inference with existing approaches requires that auxiliary information, such as known population means or totals, is available to survey data users.

In practice, public-use survey data are released to users and such data sets often report only the variables of interest and the final survey weights $\{w_i, i\in {\mS}\}$ obtained by adjusting for unit nonresponse and calibration on auxiliary variables selected by the producer of the data, where $\mS$ denotes the set of units included in the released data file. Furthermore, the data file provides $B$ columns of final replication weights $\{w_i^{(b)}, i\in {\mS}\}$ designed for variance estimation. The following table shows a typical format of public-use survey data files as seen by the users.

\medskip

\begin{table}[htpb]
\begin{center}
\begin{tabular}{| c | c c c c c | c | c c c |}
\hline
\multicolumn{1}{|c|}{$\;\,$ $i$ $\;\,$}
&\multicolumn{1}{c}{$\;\,$ $y_{i1}$ $\;$}
&\multicolumn{1}{c}{$\;$ $y_{i2}$ $\;$}
&\multicolumn{1}{c}{$\;$ $x_{i1}$ $\;$}
&\multicolumn{1}{c}{$\;$ $x_{i2}$ $\;$}
&\multicolumn{1}{c}{$\;$ $x_{i3}$ $\;\,$}
&\multicolumn{1}{|c|}{$\;\,$ $w_i$ $\;\,$}
&\multicolumn{1}{c}{$\;\,$ $w_i^{(1)}$ $\;$}
&\multicolumn{1}{c}{$\cdots$}
&\multicolumn{1}{c|}{$\;$ $w_i^{(B)}$ $\;\,$}\\ [+2pt]
\hline
$1$     & $y_{11}$ & $y_{12}$ & $x_{11}$ & $x_{12}$ & $x_{13}$ & $w_1$ & $w_1^{(1)}$ & $\cdots$ & $w_1^{(B)}$\\
$2$     & $y_{21}$ & $y_{22}$ & $x_{21}$ & $x_{22}$ & $x_{23}$ & $w_2$ & $w_2^{(1)}$ & $\cdots$ & $w_2^{(B)}$\\
$3$     & $y_{31}$ & $y_{32}$ & $x_{31}$ & $x_{32}$ & $x_{33}$ & $w_3$ & $w_3^{(1)}$ & $\cdots$ & $w_3^{(B)}$\\
$\vdots$ & $\vdots$ & $\vdots$ & $\vdots$ & $\vdots$ & $\vdots$ & $\vdots$ & $\vdots$ & $\vdots$ & $\vdots$\\
$n$   & $y_{n1}$ & $y_{n2}$ & $x_{n1}$ & $x_{n2}$ & $x_{n3}$ & $w_n$  & $w_n^{(1)}$ & $\cdots$ & $w_n^{(B)}$\\
\hline
\end{tabular}
\end{center}
\end{table}

The final replication weights $\{w_i^{(b)}, i\in {\mS}\}$ are one of the most crucial parts in creating public-use survey data files. Different versions of bootstrap replication weights, such as those developed by Rao and Wu (1988) and Rao, Wu and Yue (1992) for stratified multi-stage designs, are commonly reported with the data file. Final replication weights are typically obtained by subjecting the basic replication weights (such as the bootstrap weights) to the same unit nonresponse adjustment and calibration procedures. None of the existing empirical likelihood methods is applicable for statistical inferences  with public-use data files because the first order inclusion probabilities, the calibration variables and the associated known population means or totals are not reported on the data file and are not available to users.

The main purpose of this article is to develop empirical likelihood methods for statistical analysis with public-use survey data files. We consider two general approaches: the first is based on the pseudo empirical likelihood and the second uses the sample empirical likelihood. We present design-based inferential procedures and theoretical results on two general statistical inference problems with the vector of finite population parameters defined through the census estimating equations: the maximum empirical likelihood estimators and the empirical likelihood ratio test on a general linear or nonlinear hypothesis.  Design-based variable selection through a penalized pseudo or sample empirical likelihood is discussed. We also present a bootstrap procedure under single stage survey designs for creating valid replication weights with theoretical justifications. Simulation results and an application to the General Social Survey 2016 public-use data file released by Statistics Canada are included. 

The basic settings are described in Section 2. Main theoretical results are presented in Section 3. A bootstrap procedure under single stage survey designs to create valid replication weights is described in Section 4 with theoretical justification given in the Appendix. Results from simulation studies are reported in Section 5. The application to the General Social Survey 2016 public-use data file is presented in Section 6.  We conclude with some additional remarks in Section 7. 
Our presentation on sample empirical likelihood follows Zhao, Haziza and Wu (2018), and the discussion on pseudo empirical likelihood follows Zhao and Wu (2019). Proofs and technical details of several main theoretical results have similarities to Qin and Lawless (1994), Zhao, Haziza and Wu (2018) and Zhao and Wu (2019), and are presented in the Appendix.

\medskip

\setcounter{section}{2}
\setcounter{equation}{0}
\no {\bf 2. Empirical Likelihood and Estimating Equations for Complex Surveys}

\medskip

\no
Let \ ${\mU} = \{1,2,\cdots,N\}$ be the set of units in the finite population, where $N$ is the population size. Let $(y_i,x_i)$ be the measures of the study variable $y$ and auxiliary variables $x$ for unit $i$. Let ${\mF}_{\N} = \{(y_i,x_i), i=1,\cdots,N\}$ represent the survey population and let $\{(y_i,x_i), i\in {\mS}\}$ be the survey sample data. Let $\pi_i = P(i\in \mS)$, $i=1,\cdots, N$ be the first order inclusion probabilities.

Survey data are a major source of information for official statistics, where the focus is often on descriptive population quantities such as population means or quantiles. Complex surveys are also frequently used by researchers in social sciences and medical and health studies for statistical modelling. Under both scenarios, the finite population parameters $\theta_{\N}$ of dimension $p$ can be defined as the solution to the census estimating equations
\begin{equation}
U_{\N}(\theta) = \sum_{i=1}^N g(x_i,y_i,\theta) = 0 \,,
\label{UN}
\end{equation}
where $g(x,y,\theta)$ is an estimating function of dimension $r$, and $\theta\in \Theta$, a compact subset of $\mathcal {R}^{p}$ with $1\leq p \leq r$. Under normal circumstances we have $r=p$ but over-identified scenarios with $r>p$ do arise in practice due to additional calibration constraints or known moment conditions over certain variables.

Standard empirical likelihood inference with independent observations, as introduced by Owen (1988) and with parameters defined by estimating equations, as discussed by Qin and Lawless (1994), consists of three ingredients:
\begin{eqnarray}
&& \ell(\bfp) = \sum_{i\in \mS}\log(p_i) \,, \label{ELfunction0} \\
&& \sum_{i\in \mS}p_i = 1 \,, \label{norm0} \\
&& \sum_{i\in \mS} p_i g(x_i,y_i,\theta) = 0 \,, \label{para0}
\end{eqnarray}
where $\ell(\bfp)$ given by (\ref{ELfunction0}) is the empirical log-likelihood function and $\bfp=(p_1,\cdots,p_n)$ is the probability measure over the $n$ sampled units, equation (\ref{norm0}) is the normalization constraint to ensure that $\bfp$ is a discrete probability measure, and equations (\ref{para0}) are the constraints induced by the parameters $\theta$. The use of $\log(p_i)$ implicitly requires that $p_i >0$.

Naive applications of the standard empirical likelihood methods to complex survey data do not produce valid results under the design-based framework. There have been three major modified approaches in the survey sampling literature on using the empirical likelihood method for complex survey data, and their relations to the standard empirical likelihood ingredients (\ref{ELfunction0}), (\ref{norm0}) and (\ref{para0}) can be described as follows.

\medskip

\noindent
(1) {\em The pseudo empirical likelihood approach (PEL)}: Chen and Sitter (1999) suggested to replace $\ell(\bfp)$ by
$\ell_{\PELz}(\bfp) = \sum_{i \in \mS}d_i \log(p_i)$, where $d_i = \pi_i^{-1}$ are the basic design weights, while constraints (\ref{norm0}) and (\ref{para0}) remain unchanged. The use of $\ell_{\PELz}(\bfp)$ is motivated by the fact that $\ell_{\PELz}(\bfp)$ is the Horvitz-Thompson estimator for the conceptual census empirical log-likelihood function $\sum_{i=1}^N \log(p_i)$. Wu and Rao (2006) used a modified version $\ell_{\PELo}(\bfp) = n \sum_{i \in \mS}\tilde{d}_i(\mS) \log(p_i)$, where $\tilde{d}_i(\mS) = d_i / \sum_{j\in \mS} d_j$, which facilitates the construction of the pseudo empirical likelihood ratio confidence intervals for population parameters.  Rao and Wu (2010a) extended the method for multiple frame surveys and Rao and Wu (2010b) developed a Bayesian pseudo empirical likelihood method to survey data analysis. However, all the existing results on pseudo empirical likelihood methods focus primarily on inferences for a scalar parameter. General statistical tools involving a vector of parameters with the pseudo empirical likelihood are not available.

\medskip

\noindent
(2) {\em The population empirical likelihood approach (POEL)}: Chen and Kim (2014) defined the population empirical log-likelihood function as
$\ell_{\POEL} = \sum_{i=1}^N\log(\omega_i)$ with normalization constraint $\sum_{i=1}^N \omega_i =1$. The survey data and parameters are forced into the ``population system'' through the constraints $\sum_{i\in \mS} \omega_i \pi_i^{-1} =1$ and $\sum_{i\in \mS} \omega_i \{g(x_i,y_i,\theta) \pi_i^{-1}\}=0$. Chen and Kim (2014) focused on Poisson sampling and rejective sampling, and the method has not been developed for general unequal probability sampling designs or general inferential problems for analytical use of survey data.

\medskip

\noindent
(3) {\em The sample empirical likelihood approach (SEL)}: The method was first mentioned very briefly by Chen and Kim (2014) as a remark but detailed exploration was not pursued in their paper. The idea is to use the standard empirical log-likelihood function $\ell_{\SELz}(\bfp) = \sum_{i\in \mS}\log(p_i)$ from (\ref{ELfunction0}) and the standard normalization constraint (\ref{norm0}) but modify the constraints induced by the parameters as
$\sum_{i\in \mS} p_i \{g(x_i,y_i,\theta) \pi_i^{-1}\}=0$. A related formulation was presented by Berger and De La Riva Torres (2016) and Oguz-Alper and Berger (2016). They used $l_{(m)} = \sum_{i\in \mS} \log(m_i)$, where the $m_i$ satisfy the so-called design constraint $\sum_{i\in \mS} m_i \pi_i = n$. 
The constraints for the parameters are specified as $\sum_{i\in \mS} m_i g(x_i,y_i,\theta) = 0$. It can be seen that, if we let $p_i = m_i \pi_i n^{-1}$, the formulation is equivalent to the one proposed by Chen and Kim (2014). The sample empirical likelihood method has been further developed in a recent paper by Zhao, Haziza and Wu (2018) as a general inference tool for survey data analysis under the assumption that the first order inclusion probabilities $\pi_i$ and other related design and population information are available.

\medskip

Unfortunately, none of the existing empirical likelihood methods can be used directly for statistical analysis with public-use survey data files since the initial inclusion probabilities $\pi_i$ are not available, and calibration variables along with their known population totals are typically not given to the end users of the data files. On the other hand, the availability of final survey weights and replication weights for public-use data sets provides a unique opportunity to develop empirical likelihood as a general statistical tool for survey data analysis.

\medskip

\setcounter{section}{3}
\setcounter{equation}{0}
\no {\bf 3. Empirical Likelihood Inference with Public-Use Survey Data}

\medskip

\no {\bf 3.1  Public-use survey data and basic assumptions}

\medskip

\no
Consider the following version of a micro survey data file, which is released by the survey agency for public use:
\[
\Bigl\{\Bigl(y_i, x_i, w_i, w_i^{(1)}, \ldots, w_i^{(B)}\Bigr),\; i=1,2,\ldots,n\Bigr\} \,,
\]
where the $y_i$ and $x_i$ are possibly vector-values survey variables included in the data set, the $w_i$ is the final survey weight for unit $i$ after unit nonresponse adjustment and/or calibration weighting, and $n$ is the final sample size. Also included in the data file are $B$ final replication weights $w_i^{(1)}$, $\ldots$, $w_i^{(B)}$ associated with unit $i$. The detailed survey design information such as the original design weights $d_i = 1/\pi_i$ and the known auxiliary population information are assumed to be unavailable to the users of the data file. It is also assumed that the finite population size $N$ is unknown.

The survey weighted estimating equations for the vector of parameters $\theta_{\N}$ defined by the census estimating equations (\ref{UN}) are given by
\begin{equation}
\hat{U}_n(\theta) = \sum_{i\in \mS} w_i \, g(x_i,y_i,\theta) = 0\,.
\label{Un}
\end{equation}
For standard scenarios where $r=p$, i.e., the number of equations is the same as the number of parameters, the survey weighted estimator $\hat\theta_{\N}$ for $\theta_{\N}$ is the solution to (\ref{Un}). Let $g_i(\theta) = g(x_i,y_i,\theta)$ and assume that $g_i(\theta)$ is a smooth function of $\theta$. The approximate design-based variance of $\hat\theta_{\N}$ has the well-known sandwich form (Binder, 1983)
\[
Var\big(\hat\theta_{\N}\big) \doteq \Gamma^{-1} Var\big\{N^{-1}\hat{U}_n(\theta_{\N})\big\} \big(\Gamma^{-1}\big)' \,,
\]
where $\Gamma = \Gamma(\theta_{\N})$, $\Gamma(\theta) = N^{-1}\sum_{i=1}^N \partial g_i(\theta)/\partial \theta$ and $Var\big\{N^{-1}\hat{U}_n(\theta_{\N})\big\}$ is the design-based variance. There have been attempts to address hypothesis testing problems involving a single component of the vector of parameters $\theta_{\N}$ under the estimating equations framework, see, for instance, Binder and Patak (1994), but general hypothesis testing procedures are not available in the literature.

We consider smooth estimating functions and allow over-identified estimating equations system with $r\ge p$. Practically useful results for the special case $r=p$ and for a scalar parameter (i.e., $p=1$) will also be spelled out.
For asymptotic development, we assume that there is a sequence of finite populations and a sequence of survey designs with both the population size $N$ and the sample size $n$ going to infinity; see Isaki and Fuller (1982) for further detail. We use $N \rightarrow \infty$ to denote the limiting process. Note that $\theta_{\N}$ refers to the true vector of the finite population parameters. Throughout the paper, we use $\|\cdot\|$ to denote  the Euclidean norm and $\stackrel{{\cal L}}{\rightarrow} $ to denote convergence in distribution under the  design-based framework. Let
$O_p(\cdot)$ and $o_p(\cdot)$ be the stochastic orders under the same framework. We consider the following basic assumptions for the public-use survey data file and the estimating functions $g_i(\theta)$.

\medskip

\noindent 
{\bf Assumption 1.} {\em 
The final survey weights $(w_1, w_2, \ldots, w_n)$ and the finite population values ${\mF}_{\N} = \{(y_i,x_i), i=1,\cdots,N\}$ satisfy conditions that ensure 
$\hat{U}_n(\theta_{\N}) = \sum_{i\in \mS} w_ig_i(\theta_{\N})$
is asymptotically normally distributed with mean zero and variance-covariance matrix of the order $O(N^2/n)$.
}

\medskip

Let $\hat{\eta}^{(b)}(\theta_{\N}) = \sum_{i\in \mS} w_i^{(b)} g_i(\theta_{\N})$ be the replicate version of $\hat{U}_n(\theta_{\N}) = \sum_{i\in \mS} w_ig_i(\theta_{\N})$ using the $b$th set of replication weights $(w_1^{(b)}, w_2^{(b)}, \ldots, w_n^{(b)})$, $b=1,2,\ldots,B$ and treating $\theta_{\N}$ as a known number.

\medskip

\noindent 
{\bf Assumption 2.} {\em 
The final replication weights ensure that the replication variance estimator
\begin{equation}
v\bigl\{ \hat{U}_n(\theta_{\N})\bigr\} = \frac{1}{B} \sum_{b=1}^B \Bigl\{\hat{\eta}^{(b)}(\theta_{\N}) - \hat{U}_n(\theta_{\N}) \Bigr\}\Bigl\{\hat{\eta}^{(b)}(\theta_{\N}) - \hat{U}_n(\theta_{\N}) \Bigr\}'
\label{vU}
\end{equation}
is a design-consistent estimator of the variance-covariance matrix $Var\bigl\{\hat{U}_n(\theta_{\N})\mid {\mF}_{\N}\bigr\}$.
}

\medskip

The original design weights, the nonresponse adjusted weights and the calibration weights usually satisfy Assumption 1. It is part of the foundation for design-based inference. 
Assumption 2 is the guiding principle for public-use data file producers on how to create replication weights and for research activities on replication methods for variance estimation in surveys. Note that Assumption 2 does not necessarily require a large $B$ for the given data set, as shown by the results presented in Kim and Wu (2013). Most survey organizations, including Statistics Canada, use $B=500$ for producing public-use survey data files in their current practice. See the example of General Social Survey presented in Section 6.

\medskip

\noindent 
{\bf Assumption 3.} {\em 
(i) $\lim_{N\rightarrow \infty}(n/N)=\gamma\in(0,1)$; (ii) $c_1 < w_i N / n < c_2$, $i\in \mS$ for some positive constants $c_1$ and $c_2$;  (iii) $N^{-1}\sum_{i\in \mS}w_i-1=O_p(n^{-1/2})$.
}

\medskip

\noindent
{\bf Assumption 4.} {\em
(i)
    $\sup_{\theta\in\Theta} N^{-1}\sum_{i\in \mS}\|g_i(\theta)\|^{\kappa}<c$ for some $\kappa > 2$ and some positive constant $c$; (ii)
$\max_{i\in \mS}\sup_{\theta\in\Theta}\|g_i(\theta)\|=o_p(n^{1/2})$.
}

\medskip

\noindent 
{\bf Assumption 5.} {\em 
(i)
The matrices $W_1(\theta_{\N})=N^{-1}\sum_{i=1}^N g_i(\theta_{\N})g_i(\theta_{\N})'$, $W_2(\theta_{\N})= nN^{-2} $ $E\{\sum_{i\in \mS} w_i^2g_i(\theta_{\N})g_i(\theta_{\N})' \mid {\mF}_{\N}\}$ and $\Omega(\theta_{\N})= n N^{-2}Var\{\sum_{i\in \mS}w_ig_i(\theta_{\N})\mid {\mF}_{\N}\}$ are all positive definite;
(ii)
$\Gamma(\theta_{\N})=N^{-1}\sum_{i=1}^N\partial g_i(\theta)/\partial\theta |_{\theta=\theta_{\N}}$ has full column rank $p$.
}

\medskip

Assumptions 3-5 are standard regularity conditions for asymptotic development for finite populations with complex survey data.
The inclusion of the factors $N^{-1}$ or $N^{-2}$ in the quantities presented in Assumption 5 is for convenience in asymptotic orders. They are not required for computational purposes as they all cancel out in the main results to be presented in the next two subsections.
The pseudo empirical likelihood approach of Section 3.2 and the sample empirical likelihood approach of Section 3.3 are formulated using the final weights $w_i$. The empirical likelihood ratio statistics for both approaches do not have standard $\chi^2$ asymptotic distributions, since design-based variances require information from additional columns of replication weights in the dataset.

\medskip

\no {\bf 3.2  The pseudo empirical likelihood approach}

\medskip

\no
Let $\tilde{w}_i({\mS}) = w_i /\sum_{k\in {\mS}}w_k$, $i\in {\mS}$ be the normalized final survey weights. The pseudo empirical log-likelihood function is defined as
\[
\ell_{\PEL}(\bfp) = n \sum_{i\in {\mS}}\tilde{w}_i({\mS}) \, \log(p_i) \,.
\]
For the special case of equal final survey weights, we have $\tilde{w}_i({\mS}) = 1/n$ and $\ell_{\PEL}(\bfp) = \sum_{i\in {\mS}} \log(p_i)$. Maximizing $\ell_{\PEL}(\bfp)$ subject to the normalization constraint (\ref{norm0}), i.e., $\sum_{i\in {\mS}}p_i $ $= 1$, gives $\hat{\bfp} = (\hat{p}_1,\ldots,\hat{p}_n)$, where $\hat{p}_i = \tilde{w}_i({\mS})$. Let $\hat{\bfp}(\theta) = (\hat{p}_1(\theta),\ldots,\hat{p}_n(\theta))$ be the maximizer of $\ell_{\PEL}(\bfp)$ under the normalization constraint (\ref{norm0}) and the parameter constraint (\ref{para0}), i.e., $\sum_{i\in {\mS}} p_i \, g_i(\theta) = 0$,
for a fixed value of $\theta$. It can be shown that
$\hat{p}_i(\theta) = \tilde{w}_i({\mS})/\{1+\lambda' g_i(\theta)\}$
for $i\in \mS$, where the Lagrange multiplier $\lambda=\lambda(\theta)$ is the solution to
\begin{equation}
g_{\PEL}(\lambda) = \sum_{i\in {\mS}} \frac{\tilde{w}_i({\mS}) g_i(\theta)}{1+\lambda' g_i(\theta)} = 0 \,,
\label{lam1}
\end{equation}
which can be solved using the modified Newton-Raphson method presented in Chen, Sitter and Wu (2002) and the R code described in Wu (2005). The maximum pseudo empirical likelihood estimator $\hat{\theta}_{\PEL}$ is the maximizer of
$\ell_{\PEL}\bigl\{\hat{\bfp}(\theta)\bigr\} = n \sum_{i\in {\mS}} \tilde{w}_i({\mS}) \log\bigl\{\hat{p}_i(\theta)\bigr\}$
with respect to $\theta$. For the special case $r=p$, the estimator $\hat{\theta}_{\PEL}$ is the solution to
\[
\sum_{i\in {\mS}} \hat{p}_i \, g_i(\theta) = \sum_{i\in {\mS}} \tilde{w}_i({\mS}) \, g_i(\theta) = 0\,,
\]
which is the same as the customary survey weighted estimating equations estimator $\hat\theta_{\N}$.
The pseudo empirical log-likelihood ratio statistic for $\theta$ is given by
\[
r_{\PEL}(\theta) = \ell_{\PEL}\bigl\{\hat{\bfp}(\theta)\bigr\} - \ell_{\PEL}\bigl(\hat{\bfp}\bigr)
 =  - n \sum_{i\in \mS} \tilde{w}_i(\mS) \log\bigl\{1+\lambda' g_i(\theta)\bigr\} \,.
\]
We can re-write the maximum pseudo empirical likelihood estimator of $\theta_{\N}$ as
$\hat{\theta}_{\PEL}=\mathop{\arg\max}_{\theta\in\Theta}r_{\PEL}(\theta)$.
The following theorem presents asymptotic properties of the estimator $\hat{\theta}_{\PEL}$. Note that the quantities $W_1(\theta_{\N})$, $\Gamma(\theta_{\N})$ and $\Omega(\theta_{\N})$ are defined in Assumption 5.

\medskip

\noindent
{\bf Theorem 1.} {\em 
Under Assumptions  1, 3, 4 and 5, we have
\[
n^{1/2}(\hat{\theta}_{\PEL}-\theta_{\N})\mid {\mF}_{\N}\;\; \stackrel{{\cal L}}{\longrightarrow} \;\; N(0, V_1) \,,
\]
where  $V_1=\Sigma_1\Gamma'W_1^{-1}\Omega W_1^{-1}\Gamma \Sigma_1$,
$\Sigma_1=(\Gamma' W_1^{-1}\Gamma)^{-1}$, $W_1=W_1(\theta_{\N})$, $\Gamma=\Gamma(\theta_{\N})$ and $\Omega=\Omega(\theta_{\N})$.
}

\medskip

Proofs of Theorem 1 and Theorems 2-6 presented below resemble the proofs in Zhao et al. (2018). Details are presented in the Appendix. The proof of Theorem 1 is also similar to the proof of Theorem 1 in Qin and Lawless (1994). 

\medskip

\noindent
{\bf Corollary 1.} {\em 
Under the assumptions in Theorem 1 and $r=p$ (i.e., the number of equations is the same as the number of parameters), the asymptotic variance-covariance matrix $V_1$ for $\hat{\theta}_{\PEL}$ reduces to $V_{1} = \Gamma^{-1} \Omega (\Gamma')^{-1}$.
}

\medskip

Suppose we want to test the simple hypothesis: $H_0: \theta_{\N}=\theta_{\Nzero}$ against $H_1: \theta_{\N} \neq \theta_{\Nzero}$. The pseudo empirical log-likelihood ratio statistic for testing $H_0$ is given by 
\[
{\rm LR}_{\PEL}(\theta_{\Nzero}) = 2\big\{r_{\PEL}(\hat\theta_{\PEL})-r_{\PEL}(\theta_{\Nzero})\big\}
                                              = 2\big\{\ell_{\PEL}(\hat\theta_{\PEL})-\ell_{\PEL}(\theta_{\Nzero})\big\}\,.
\]
The asymptotic distribution of ${\rm LR}_{\PEL}(\theta_{\Nzero})$ is given by the following theorem.

\medskip

\noindent
{\bf Theorem 2.} {\em 
Suppose that Assumptions  1, 3, 4 and 5 hold.  Then
\[
{\rm LR}_{\PEL}(\theta_{\Nzero})\mid {\mF}_{\N} \;\; \stackrel{{\cal L}}{\longrightarrow} \;\; Q'\Delta_1Q \,,
\]
where $Q \sim N(0, I_r)$, $I_r$ is the $r\times r$ identity matrix, $r$ is the dimension of the estimating functions $g_i(\theta)$,
and
$
\Delta_1 = \Omega^{1/2}W_1^{-1}\Gamma\Sigma_1  \Gamma'W_1^{-1}\Omega^{1/2}
$
with $\Sigma_1=(\Gamma' W_1^{-1}\Gamma)^{-1}$.
}

\medskip

\noindent
{\bf Corollary 2.} {\em 
Under the assumptions in Theorem 2 and $r=p$, we have $\Delta_1 = \Omega^{1/2}W_1^{-1}\Omega^{1/2}$.
In particular, if $r=p=1$, then
\[
{\rm LR}_{\PEL}(\theta_{\Nzero})\mid {\mF}_{\N} \;\; \stackrel{{\cal L}}{\longrightarrow} \;\; (\Omega/W_1)\chi^2(1) \,,
\]
where $\chi^2(1)$ denotes  the standard $\chi^2$ random variable with one degree of freedom.
}

\medskip

We further consider pseudo empirical log-likelihood ratio test for a general linear or nonlinear hypothesis $H_0$: $R(\theta_{\N}) = 0$ against $H_1$: $R(\theta_{\N}) \neq 0$, where $R(\theta_{\N})$ is a  $k\times 1$ vector-valued functions with  $k\leq p$ and $R(\theta_{\N}) = 0$ imposes $k$ constraints on the vector of parameters $\theta_{\N}$. Let $\Theta^{\R} = \big\{ \theta \mid \theta \in \Theta \; {\rm and} \; R(\theta) = 0\big\}$
be the restricted parameter space under $H_0$. The restricted maximum pseudo empirical
likelihood estimator of $\theta$ under $H_0$ is defined as
$\hat\theta_{\PEL}^{\R}=\mathop{\arg\max}_{\theta\in \Theta^{\R}}r_{\PEL}(\theta).$  The  pseudo empirical log-likelihood ratio statistic for testing $H_0$ versus $H_1$ is given by
\[
{\rm LR}_{\PEL}(\theta_{\N} \mid H_0) = 2\big\{r_{\PEL}(\hat\theta_{\PEL})-r_{\PEL}(\hat\theta_{\PEL}^{\R})\big\}
= 2\big\{\ell_{\PEL}(\hat\theta_{\PEL})-\ell_{\PEL}(\hat\theta_{\PEL}^{\R})\big\}\,.
\]

\medskip

\noindent
{\bf Theorem 3.} {\em 
Suppose that Assumptions  1, 3, 4 and 5 hold.
If the function $R(\theta)$ is twice continuously differentiable and $\Phi(\theta_{\N})=\partial R(\theta)/\partial\theta |_{\theta=\theta_{\N}}$ has rank $k$, then
\[
{\rm LR}_{\PEL}(\theta_{\N} \mid H_0) \mid {\mF}_{\N} \;\; \stackrel{{\cal L}}{\longrightarrow} \;\; Q'\Delta_1^{\R}Q \,,
\]
where \  $Q \sim N(0, I_r)$, \
$\Delta_1^{\R} = \Omega^{1/2}W_1^{-1}\Gamma\Sigma_1 \Phi'(\Phi\Sigma_1 \Phi)^{-1}\Phi\Sigma_1\Gamma'W_1^{-1}\Omega^{1/2}$ \ and \ $\Phi=\Phi(\theta_{\N})$.
}

\medskip

Let $\delta_j$, $j=1,\cdots,p$ be the non-zero eigenvalues of the $r\times r$ matrix $\Delta_1$.
The asymptotic distribution of ${\rm LR}_{\PEL}(\theta_{\N})$ given in Theorem 2 can be alternatively represented by $\sum_{j=1}^p  \delta_j \chi^2_j(1)$, where $\chi^2_j(1)$, $j=1,\cdots,p$ are independent random variables, all following the same distribution as $\chi^2(1)$. Similarly, the distribution of the quadratic form $Q^{\T}\Delta_1^{\R}Q$ given in Theorem 3 can be alternatively represented by $\sum_{j=1}^k  \delta_j^{\R} \chi^2_j(1)$, where $\delta_j^{\R}$, $j=1,\cdots,k$ are the non-zero eigenvalues of the matrix $\Delta_1^{\R}$.

Practical implementations of the theoretical results generally require the estimation of the asymptotic variance $V_1$ for Theorem 1, the matrix $\Delta_1$ for Theorem 2 and $\Delta_1^{\R}$ for Theorem 3. This amounts to estimating the involved components $W_1$, $\Gamma$, $\Omega$ and $\Phi$.
By the simple ``plug-in" method, we can estimate the term $W_1$ by
$\hat{W}_1=N^{-1}\sum_{i\in \mS} w_ig_i(\hat{\theta}_{\PEL})g_i(\hat{\theta}_{\PEL})'$,
the term $\Gamma$ by
$\hat{\Gamma}_{\PEL}=N^{-1}\sum_{i\in \mS} w_i\partial g_i(\theta)/\partial\theta |_{\theta=\hat{\theta}_{\PEL}}$, and estimate $\Phi=\Phi(\theta_{\N})$ by
$\hat\Phi=\Phi(\hat\theta_{\PEL})$.
The most critical component $\Omega$ can be estimated by  $\hat{\Omega}_{\PEL}=nN^{-2}v\bigl\{ \hat{U}_n(\hat{\theta}_{\PEL})\bigr\}$, where
$v\bigl\{ \hat{U}_n(\hat{\theta}_{\PEL})\bigr\}$ is the replication variance estimator outlined in Assumption 2 using the replication weights from the survey data file.

The distribution of the quadratic forms $Q'\Delta_1Q$ and $Q'\Delta_1^{\R}Q$ may also be approximated by the Rao-Scott (RS) correction method (Rao and Scott, 1981, 1984). For instance,
the first-order RS correction leads to ${\rm LR}_{\PEL}(\theta_{\N})\mid {\mF}_{\N}  \;\; \stackrel{{\cal L}}{\longrightarrow} \;\; a\chi^{2}(p)$, where $a=\sum_{j=1}^{p}\delta_j/p$. The second-order
RS correction gives ${\rm LR}_{\PEL}(\theta_{\N})\mid {\mF}_{\N}  \;\; \stackrel{{\cal L}}{\longrightarrow} \;\; c\chi^{2}(k^*)$, where $c=\sum_{j=1}^{p}\delta_j^2/\sum_{j=1}^{p}\delta_j$ and $k^*=(\sum_{j=1}^{p}\delta_j)^2/\sum_{j=1}^{p}\delta_j^2$.

\medskip

\no {\bf 3.3  The sample empirical likelihood approach}

\medskip

\no
The sample empirical likelihood approach described in \S 2 can be adapted for public-use survey data. We start with the standard empirical log-likelihood function $\ell_{\SEL}(\bfp) = \sum_{i\in \mS} \log(p_i)$.
Maximizing $\ell_{\SEL}(\bfp)$ under the normalization constraint (\ref{norm0}), i.e., $\sum_{i\in \mS} p_i = 1$, gives $\hat{p}_i = n^{-1}$, $i\in \mS$. The constraints for the parameters $\theta$ defined through (\ref{UN}) are formed using the weighted estimating functions $w_ig_i(\theta)$ and are given by
\begin{equation}
\sum_{i\in \mS} p_i \bigl\{w_i g_i(\theta)\bigr\} = 0\,.
\label{wtheta}
\end{equation}
Let $\hat{\bfp}(\theta) = (\hat{p}_1(\theta),\ldots,\hat{p}_n(\theta))$ be the maximizer of $\ell_{\SEL}(\bfp)$ under the normalization constraint (\ref{norm0}) and the parameter constraints (\ref{wtheta}) for a fixed $\theta$. It follows from standard empirical likelihood method that
$\hat{p}_i(\theta) = n^{-1} [1+\lambda' \{w_i g_i(\theta)\}]^{-1}$
for $i \in \mS$, where the Lagrange multiplier $\lambda=\lambda(\theta)$ is the solution to 
\begin{equation}
g_{\SEL}(\lambda) = \frac{1}{n}\sum_{i\in \mS} \frac{w_ig_i(\theta)}{1+\lambda' \{w_i g_i(\theta)\}} = 0 \,.
\label{lam2}
\end{equation}
The empirical log-likelihood ratio statistic for $\theta$ under the current setting is given by
\[
r_{\SEL}(\theta) = \ell_{\SEL}\bigl\{\hat{\bfp}(\theta)\bigr\} - \ell_{\SEL}\bigl(\hat{\bfp}\bigr)
= \sum_{i\in \mS} \log\{n\hat{p}_i(\theta)\}
= - \sum_{i\in \mS} \log\{1+\lambda' w_i g_i(\theta)\}\,.
\]
Let $\hat{\theta}_{\SEL}=\mathop{\arg\max}_{\theta\in\Theta}r_{\SEL}(\theta)$ be the maximum sample empirical likelihood estimator of $\theta_{\N}$. We have the following major results on the asymptotic properties of $\hat{\theta}_{\SEL}$.

\medskip

\noindent
{\bf Theorem 4.} {\em 
Suppose that Assumptions  1, 3, 4 and 5 hold. Then
\[
n^{1/2}(\hat{\theta}_{\SEL}-\theta_{\N})\mid {\mF}_{\N} \;\; \stackrel{{\cal L}}{\longrightarrow}\;\; N(0, V_2) \,,
\]
where  $V_2=\Sigma_2\Gamma' W_2^{-1}\Omega W_2^{-1}\Gamma \Sigma_2$ with
 $\Sigma_2=(\Gamma' W_2^{-1}\Gamma)^{-1}$.
}

\medskip

The results presented in Theorem 4 under the sample empirical likelihood are similar to those in Theorem 1 for the pseudo empirical likelihood, with the crucial differences in defining $W_1$ for Theorem 1 and $W_2$ in Theorem 4. For the special case $r=p$, the estimator $\hat{\theta}_{\SEL}$ is attained as the global maximum point with $\hat{p}_i=n^{-1}$ and is the solution to  $\sum_{i\in \mS}w_ig_i(\theta)=0$, which coincides with the survey weighted estimating equations estimator.

\medskip

\noindent
{\bf Corollary 3.} {\em 
Suppose that the assumptions of Theorem 4 hold.
If $r=p$, then the asymptotic variance-covariance matrix $V_2$ for $\hat{\theta}_{\SEL}$ reduces to $V_{2} = \Gamma^{-1} \Omega (\Gamma')^{-1}$.
}

\medskip

The sample empirical log-likelihood ratio statistic for testing $H_0: \theta=\theta_{\N}$ is similarly defined as
\[
{\rm LR}_{\SEL}(\theta) = 2\big\{r_{\SEL}(\hat\theta_{\SEL})-r_{\SEL}(\theta)\big\}
                                              = 2\big\{\ell_{\SEL}(\hat\theta_{\SEL})-\ell_{\SEL}(\theta)\big\}
\]
for the given $\theta$. We have the following results parallel to Theorem 2 and Corollary 2. Once again, the differences are between $W_1$ and $W_2$ involved in the asymptotic distributions.

\medskip

\noindent
{\bf Theorem 5.} {\em 
Suppose that Assumptions  1, 3, 4 and 5 hold.  Then
\[\begin{array}{lllll}
{\rm LR}_{\SEL}(\theta_{\N})\mid {\mF}_{\N} \;\; \stackrel{{\cal L}}{\longrightarrow} \;\; Q'\Delta_2Q \,,
\end{array}
\]
where $Q \sim N(0, I_r)$ and $\Delta_2 = \Omega^{1/2}W_2^{-1}\Gamma\Sigma_2  \Gamma' W_2^{-1}\Omega^{1/2}$ with $\Sigma_2=(\Gamma' W_2^{-1}\Gamma)^{-1}$.
}

\medskip

\noindent
{\bf Corollary 4.} {\em 
Suppose that the assumptions of Theorem 5 hold. If $r=p$, then
$\Delta_2 = \Omega^{1/2}W_2^{-1}\Omega^{1/2}$.
In particular, if $r=p=1$, we have ${\rm LR}_{\SEL}(\theta_{\N}) \;\; \stackrel{{\cal L}}{\longrightarrow} \;\; (\Omega/W_2)\chi^2(1)$.
}

\medskip

For a general linear or nonlinear hypothesis $H_0$: $R(\theta_{\N}) = 0$ versus $H_1$: $R(\theta_{\N}) \neq 0$, the restricted maximum sample empirical
likelihood estimator of $\theta$ under $H_0$ is defined as $\hat\theta_{\SEL}^{\R}=\mathop{\arg\max}_{\theta\in \Theta^{\R}}r_{\SEL}(\theta)$, where
$\Theta^{\R} = \big\{ \theta \mid \theta \in \Theta \; {\rm and} \; R(\theta) = 0\big\}$.   The sample  empirical log-likelihood ratio statistic for testing $H_0$  against $H_1$ is given by
\[
{\rm LR}_{\SEL}(\theta_{\N} \mid H_0) = 2\big\{r_{\SEL}(\hat\theta_{\SEL})-r_{\SEL}(\hat\theta_{\SEL}^{\R})\big\}
= 2\big\{\ell_{\SEL}(\hat\theta_{\SEL})-\ell_{\SEL}(\hat\theta_{\SEL}^{\R})\big\} \,.
\]

\medskip

\noindent
{\bf Theorem 6.} {\em 
Suppose that the assumptions of Theorem 3 hold.  If the function $R(\theta)$ is twice continuously differentiable and $\Phi(\theta_{\N})=\partial R(\theta)/\partial\theta |_{\theta=\theta_{\N}}$ has rank $k$, then
\[
{\rm LR}_{\SEL}(\theta_{\N} \mid H_0) \mid {\mF}_{\N} \;\; \stackrel{{\cal L}}{\longrightarrow} \;\; Q' \Delta_2^{\R}Q \,,
\]
where $Q \sim N(0, I_r)$, $\Delta_2^{\R} = \Omega^{1/2}W_2^{-1}\Gamma\Sigma_2 \Phi'(\Phi\Sigma_2 \Phi')^{-1}\Phi\Sigma_2\Gamma' W_2^{-1}\Omega^{1/2}$, and $\Phi=\Phi(\theta_{\N})$.
}

\medskip

The term $W_2$ for the sample empirical likelihood is different from $W_1$ for the pseudo empirical likelihood and can be estimated by
\[
\hat{W}_2=nN^{-2}\sum_{i\in \mS} w_i^2g_i(\hat{\theta}_{\SEL})g_i(\hat{\theta}_{\SEL})' \,.
\] 
The other two component $\Gamma$ and $\Omega$ can be respectively estimated by
\[
\hat{\Gamma}_{\SEL}=N^{-1}\sum_{i\in \mS} w_i\partial g_i(\theta)/\partial\theta |_{\theta=\hat{\theta}_{\SEL}} \;\;\;\;\; {\rm and} \;\;\;\;\;
\hat{\Omega}_{\SEL}=nN^{-2}v\bigl\{ \hat{U}_n(\hat{\theta}_{\SEL})\bigr\} \,,
\]
where $v\bigl\{ \hat{U}_n(\cdot)\bigr\}$ is given in Assumption 2.

\medskip

\no {\bf 3.4  Design-based variable selection}

\medskip

\no
Public-use survey data may contain observations on many variables. Variable selection is a useful technique when fitting a statistical model involving many covariates. The pseudo empirical likelihood and the sample empirical likelihood provide design-based approaches to variable selection through a penalized pseudo or sample empirical likelihood method.

Suppose that $\theta = (\theta_1,\cdots,\theta_p)'$ and $p_{\tau_n}(\cdot)$ is a pre-specified penalty function with regularization parameter $\tau_n$. The penalized pseudo empirical likelihood (PPEL) function of $\theta$ is defined as 
\[
 l_{\PPEL}(\theta) =  - n \sum_{i\in \mS} \tilde{w}_i(\mS) \log\bigl\{1+\lambda' g_i(\theta)\bigr\} - n\sum_{j=1}^p p_{\tau_n}(|\theta_j|)\,,
\]
where the Lagrange multiplier $\lambda$ solves $g_{\PEL}(\lambda) = 0$ given by (\ref{lam1}). The penalized sample empirical likelihood (PSEL) function is defined as 
\[
l_{\PSEL}(\theta) = - \sum_{i\in \mS} \log\{1+\lambda' w_i g_i(\theta)\}- n\sum_{j=1}^p p_{\tau_n}(|\theta_j|)\,,
\]
where the Lagrange multiplier $\lambda$ solves $g_{\SEL}(\lambda) = 0$ given by (\ref{lam2}).

The tuning parameter $\tau_n$ for the penalized pseudo empirical likelihood or the penalized sample empirical likelihood needs to be appropriately selected by a data-driven method. Various techniques have been proposed in the literature, including the generalized cross-validation method and the BIC method. Further details can be found in Fan and Li (2001) and Wang et al. (2007).

Let $\theta_{\N} = (\theta_{\None},\cdots,\theta_{\Np})'$ be defined by (\ref{UN}). The maximum penalized pseudo empirical likelihood estimator of $\theta_{\N}$ is defined as $\hat{\theta}_{\PPEL} = \arg\max_{\theta} l_{\PPEL}(\theta)$ and the maximum penalized sample empirical likelihood estimator of $\theta_{\N}$ is defined as $\hat{\theta}_{\PSEL} = \arg\max_{\theta} l_{\PSEL}(\theta)$. Both estimators enjoy the design-based oracle property for variable selection in the sense that ${\rm Pr}(\hat{\theta}_{\Nj} = 0 \mid \mF_{\N}) \rightarrow 1$ as $N \rightarrow \infty$ if $\theta_{\Nj} =0$, where $\hat{\theta}_{\Nj}$ is the corresponding component of $\hat{\theta}_{\PPEL}$ or $\hat{\theta}_{\PSEL}$ for estimating $\theta_{\Nj}$.

\medskip

\setcounter{section}{4}
\setcounter{equation}{0}
\no {\bf 4. Bootstrap Calibrated Empirical Likelihood Methods}

\medskip

\no
One of the most crucial features of public-use survey data files is the inclusion of replication weights. The guiding principle for the creation of replication weights is that they provide valid  results on variance estimation as outlined in Assumption 2. The major results presented in Section 3 involve the estimation of the design-based variance $\Omega$ using the replication weights, and inferential procedures are developed based on the limiting distributions presented in the theorems and corollaries.

A highly attractive approach for practical implementations of the EL-based tests is the bootstrap calibration method. The asymptotic distributions are approximated by the empirical distribution of the replicate copies of the empirical likelihood ratio statistic using the bootstrap weights. However, theoretical justifications of the bootstrap calibration method can be a challenge task and need to be developed case-by-case. In this section, we describe a bootstrap procedure for scenarios where the survey design is single-stage PPS sampling with small sampling fractions and the final survey weights are the calibration weights with known population totals of auxiliary variables. Theoretical justifications of the procedure are given in the Appendix.

Let $T_{x}$ be the known population totals for the vector $x$ of auxiliary variables used in the calibration. Let $d_i=1/\pi_i$ be the original design weights and let $\bigl\{(y_i,x_i,d_i),i\in \mS\bigr\}$ be the preliminary survey dataset. The calibration weights $w_i$ are obtained by minimizing a distance measure $D(w,d)$ between $w = (w_1,\ldots,w_n)$ and $d = (d_1,\ldots,d_n)$ subject to the calibration constraints
$\sum_{i\in \mS}w_i x_i = T_{ x}$. 
There are different distance measures available for calibration weighting. Wu and Lu (2016) contains an overview on computational algorithms and finite sample behaviours of weights from alternative calibration weighting methods. We consider the simple chisquare distance $D(w,d) = \sum_{i\in \mS} \bigl(w_i-d_i\bigr)^2/d_i$, which leads to closed form expressions for the final calibrated weights $w_i$. Let $\bigl\{(y_i,x_i,w_i),i\in \mS\bigr\}$ be the final survey dataset without replication weights.

We present bootstrap procedures for the sample empirical likelihood method on testing $H_0$: $\theta_{\N} = \theta_{\Nzero}$ against $H_1$: $\theta_{\N} \neq \theta_{\Nzero}$. The procedures are also valid for the pseudo empirical likelihood method. The proposed bootstrap procedures consist of the following steps.

\medskip

\noindent 
1. Select a bootstrap sample $\mS^*$ of size $n$ from the original sample $\mS$ using simple random sampling with replacement. Denote the bootstrap sample data by $\{(y_i, x_i, w_i),i\in\mS^*\}$. Note that $\mS^*$ may contain duplicated units from $\mS$.

\noindent
2. Compute the set of bootstrap weights $\{w_i^*,i\in \mS^*\}$ by minimizing the distance measure $\Phi(w^*,d) = \sum_{i\in \mS^*}\bigl(w_i^* - d_i\bigr)^2/d_i$ subject to the bootstrap version of the calibration constraints
$\sum_{i\in \mS^*}w_i^*  x_i = \hat{T}_{{ x}\HT}$,
where $\hat{T}_{{x}\HT} = \sum_{i\in \mS}d_i x_i$ is the Horvitz-Thompson estimator of the population totals $T_{ x}$ using the initial dataset.

\noindent
3. Define the bootstrap version of the sample empirical likelihood ratio function $r_{\SEL}(\theta)$ as
\[
r_{\SEL}^*(\theta) = - \sum_{i\in \mS^*} \log\{1+\lambda' w_i^* g_i(\theta)\}\,,
\]
where $\lambda$ is the solution to $g_{\SEL}^*(\lambda) = n^{-1}\sum_{i\in \mS^*} \{w_i^*g_i(\theta)\}/[1+\lambda' \{w_i^* g_i(\theta)\}] = 0$.
Compute the bootstrap version of the estimator
$\hat{\theta}_{\SEL}^*=\mathop{\arg\max}_{\theta\in\Theta}r_{\SEL}^*(\theta)$
and the bootstrap version of the SEL ratio  statistic
${\rm LR}_{\SEL}^*(\hat{\theta}_{\SEL}) = 2\big\{r_{\SEL}^*(\hat\theta_{\SEL}^*)-r_{\SEL}^*(\hat\theta_{\SEL})\big\}$,
where $\hat{\theta}_{\SEL}$ is the estimator obtained from the original survey dataset $\{(y_i, x_i,w_i),i\in \mS\}$.

\noindent
4. Repeat Steps 1-3 a large number $B$ times, independently, to obtain $B$ values of the bootstrap version of the SEL ratio statistic
as $\{{\rm LR}_{\SEL}^{*(1)}(\hat{\theta}_{\SEL})$, $\cdots$, ${\rm LR}_{\SEL}^{*(B)}(\hat{\theta}_{\SEL})\}$.

\medskip

Let $b_{\alpha}$ be the upper $\alpha$ quantile from the empirical distribution of the values of the bootstrap version
$\{{\rm LR}_{\SEL}^{*(1)}(\hat{\theta}_{\SEL})$, $\cdots$, ${\rm LR}_{\SEL}^{*(B)}(\hat{\theta}_{\SEL})\}$. The $\alpha$-level SEL ratio test rejects $H_0$: $\theta_{\N} = \theta_{\Nzero}$ if \ ${\rm LR}_{\SEL}(\theta_{\N}) > b_{\alpha}$.
The bootstrap calibrated $1-\alpha$ level confidence region for $\theta_{\N}$ is given by
$\mathcal {C}_{\BT} = \big\{\theta \mid {\rm LR}_{\SEL}(\theta)\leq b_{\alpha} \big\}$. 
It is shown in the Appendix that this confidence region  has correct asymptotic coverage probability.

The bootstrap procedures described above can be implemented through additional columns of replication weights to produce a public-use data file. Let $\{w_i^*, i\in \mS^*\}$ be a set of bootstrap weights described in Step 2. Let $h_i$ be the number of times that unit $i \in \mS$ is selected in $\mS^*$. Note that $0\leq h_i \leq n$ and $\sum_{i\in \mS}h_i = n$. The $b$th set of replication weights are constructed as $\{w_i^{(b)} = h_i w_i^*, i \in \mS\}$. Repeat the process for $b=1,\cdots,B$, independently, to create $B$ sets of replication weights. The bootstrap version ${\rm LR}_{\SEL}^*(\hat{\theta}_{\SEL})$ of the SEL ratio statistic can be computed by using the $(x,y)$ from the data file in conjunction with the set of replication weights.

\medskip

\setcounter{section}{5}
\setcounter{equation}{0}
\no {\bf 5. Simulation Studies}

\medskip

\no
In this section we report results from simulation studies on the finite sample performances of our proposed methods. The finite population $\{ (y_i, x_{i1}, x_{i2},x_{i3}), i=1,2,\ldots,N\}$ with size $N$ was generated from the following super population model
\[
y_i = x_i'\theta + \sigma \varepsilon_i\,, \;\;\; i=1,2,\ldots,N\,,
\]
where $\theta=(\theta_0,\theta_1,\theta_2,\theta_3)' = (1,1,1,1)'$, $x_i=(1, x_{i1}, x_{i2},x_{i3})'$,
$x_{i1} \sim$ Bernoulli $(0.5)$, $x_{i2} \sim$ Uniform$(0,1)$, $x_{i3} \sim$ $0.5$  $+$ Expomential$(2)$, and the $\varepsilon_i$'s are iid N$(0,1)$. We consider three cases for the variance $\sigma^2$ of the error terms: (i) $\sigma=\sigma_1 = 1$;
(ii) $\sigma = \sigma_2 = 3$; and (iii) $\sigma = \sigma_3 = [Var(\eta) (1 / \rho^2 - 1)]^{1/2}$ with $\eta=x'\theta$ and $\rho=0.8$. This is the controlled correlation coefficient between $y$ and the linear predictor $\eta$. 

The finite population parameters $\theta_{\N}=(\theta_{\Nzero},\theta_{\None},\theta_{\Ntwo}, \theta_{\Nthree})'$ under the linear regression model are defined as the solution to the census estimating equations $\sum_{i=1}^{N}g(x_i,y_i,\theta_{\N})=0$, where $g(x,y,\theta) = x(y - x'\theta)$. With a large $N$, the values of $\theta_{\N}$ are almost identical to the model parameters for the superpopulation.
Our simulation studies focus on examining the size and power of the proposed pseudo and sample empirical likelihood ratio tests.
We consider $\alpha$-level tests for two hypotheses: (1) $H_0$: $\theta_{\None} = 1.0$ versus $H_1$: $\theta_{\None} = b$; and (2) $H_0$: $\theta_{\None}-\theta_{\Ntwo}=0$ versus $H_1$: $(\theta_{\None},\theta_{\Ntwo}) = (b_1,b_2)$, for selected values of $b$ and $(b_1,b_2)$, 
with $\alpha = 0.05$ for both cases.

In survey practice, the process of creating the final survey weights $w_i$ and the final replication weights $w_i^{(b)}$, $b=1,2,\ldots,B$ can be very complicated. It depends on the original survey design, the scenarios for nonresponse, and the amount of known auxiliary information for calibration weighting. The replication weights often involve ad hoc approximations since many complex survey designs do not have precise bootstrap procedures or other resampling methods to produce final replication weights for general inferences. Rao and Wu (1988) and Rao, Wu and Yue (1992) contain further details on the topic. To make repeated simulation runs feasible, we consider single stage unequal probability sampling for the initial survey design, with the inclusion probabilities $\pi_i$ proportional to $x_{i3}$. The final survey weights and the final replication weights are created under two scenarios:

\medskip

\noindent
A. The final survey weights are calibrated over the known population totals of the $x_1$ and $x_2$ variables but unit nonresponse is not involved.

\medskip

\noindent
B. The final survey weights are adjusted for uniform unit nonresponse and calibrated over the known population totals of the $x_1$ and $x_2$ variables.

\medskip

For each of the two scenarios, there are two major tasks for each simulated sample: compute the final survey weights $w_i$ and create valid final replication weights $w_i^{(b)}$, $i\in \mS$. For single stage PPS sampling without replacement with a negligible sampling fraction, the with-replacement bootstrap procedures described in Section 4 produce final replication weights that satisfy Assumption 2 and are also valid for the bootstrap calibration method described in Section 4. Let $\mS_0$ be the set of initial sampled units and $n_0$ be the initial sample size under the original survey design and let $\mS$ be the set of units included in the final sample and $n$ be the final sample size.

Under Scenario A, we have $\mS=\mS_0$ and $n=n_0$ in the absence of unit nonresponse. The final weights are calibrated over the known population totals of $x_1$ and $x_2$. The replication weights are created based on the method described in Section 4.
Under Scenario B, let $d_i = 1/\pi_i$ be the initial design weights, $i\in \mS_0$. With uniform unit nonresponse, each unit in $\mS_0$ has a constant probability to be a respondent, and the final set $\mS$ of respondents has a random sample size. The unit nonresponse adjusted survey weights are computed as
\[
d_{0i} = d_i \biggl(\sum_{k \in \mS_0}d_k\biggr) / \biggl(\sum_{j \in \mS}d_j\biggr)\,, \;\; i\in \mS\,.
\]
This is the so-called ratio adjustment for uniform unit nonresponse and the adjusted survey weights satisfy \
$\sum_{i\in \mS}d_{0i} = \sum_{j\in \mS_0}d_j$.
Treating the set of adjusted weights $\{d_{0i},i\in \mS\}$ as the ``original'' design weights, the final survey weights and replication weights under the calibration constraints are created by following the same procedures used in Scenario A.

Simulation samples of size $n=400$ are selected for Scenario A from the population by the randomized systematic PPS sampling method (Goodman and Kish, 1950; Hartley and Rao, 1962). For Scenario B, initial samples of size $n_0 = 571$ are selected by the same PPS sampling method. The unit response probabilities are set to be uniform at $0.7$, resulting in final samples with expected sample size $E(n)=400$.
For both scenarios, we choose the finite population sizes as $N=20,000$ and $4,000$ such that the sampling fractions are $n/N = 2\%$ and $10\%$, the first case represents negligible sampling fractions and the second case is for non-negligible sampling fractions.
The final survey weights and the $B=500$ sets of final replication weights are created for the given scenario.

We compute the power of the PEL and SEL ratio tests for $H_0$: $\theta_{\None}=1.0$ versus $H_1$: $\theta_{\None} = b$ and for $H_0$: $\theta_{\None}=\theta_{\Ntwo}$ versus $H_1$: $(\theta_{\None},\theta_{\Ntwo})=(b_1,b_2)$ for selected values of $b$ and $(b_1,b_2)$. The power for $b=1.0$ and $(b_1,b_2)=(1.0,1.0)$ represents the size of the test, which is set at the level $0.05$. Results are based on $2,000$ simulation runs.
As a warning message for possible misuse of the PEL and SEL based tests, we first show that naively assuming the limiting distributions of the PEL and the SEL ratio tests with public survey data files as standard chisquares leads to invalid results. The sizes of the tests under Scenario A with different settings are presented in Table \ref{tab0}. It is apparent from Table \ref{tab0} that the test sizes are off by a large margin relative to the nominal value $0.05$ for all cases ranging from $0.141$ to $0.194$ for the first test and $0.186$ to $0.264$ for the second test.  

The limiting distributions of the PEL and the SEL ratio tests generally follow the distribution of a quadratic form presented in Section 3. We consider four methods to determine the critical region for each test:
I. Monte Carlo approximations to the distribution of the quadratic form using the estimated eigenvalues and the weighted $\chi^2$ distribution;
II. The first-order Rao-Scott correction method;
III. The  second-order Rao-Scott correction method;
IV. The Bootstrap calibration method as described in  Section 4. We also included a fifth method for comparisons: 
V. The Wald-test based on the point estimator $\hat\theta$ and the variance estimator $v(\hat\theta)$ for $\theta = \theta_{\None}$ or $\theta = \theta_{\None} - \theta_{\Ntwo}$ using standard normal approximation to $(\hat\theta -\theta)/\{v(\hat\theta)\}^{1/2}$. Method I uses the limiting distributions presented in Section 3. 
Methods I, II and III all require the estimation of eigenvalues of the matrix $\Delta_1$, $\Delta_1^{\R}$, $\Delta_2$ or $\Delta_2^{\R}$. The bootstrap calibration method IV is extremely time consuming for repeated simulations and the results are only included for Scenario A with $500$ simulation runs.

Tables \ref{tab1} and \ref{tab2} summarize the results on the size and power of the tests for $H_0$: $\theta_{\None} = 1.0$ versus $H_1$: $\theta_{\None} = b$ for PEL and SEL, respectively, with $n/N = 2\%$. The results for $b=1.0$ correspond to the size of the test with nominal value $0.05$ and the results for $b\neq 1.0$ represent the actual power of the test.
Tables \ref{tab3} and \ref{tab4} summarize the results on the size and power of the tests for $H_0$: $\theta_{\None}=\theta_{\Ntwo}$ versus $H_1$: $(\theta_{\None},\theta_{\Ntwo})=(b_1,b_2)$. The results for $(b_1,b_2)=(1.0,1.0)$ correspond to the size of the test and the results for other values of $(b_1,b_2)$ represent the power of the test. Simulation results corresponding to $n/N = 10\%$ are reported in the Supplementary Material. 

Major observations of the simulation results in Tables \ref{tab1}-\ref{tab4} can be summarized as follows. (1) All three approaches (i.e., PEL, SEL and  Wald)  have test sizes close to the nominal value $0.05$ for almost all cases. The PEL based tests perform the best in terms of valid test size while the SEL based tests have a few cases with sizes bigger than $0.065$. 
(2) The tests are generally more powerful when the error variance $\sigma^2$ is smaller (the cases with $\sigma_1$ and $\sigma_2$), where the auxiliary variables used for calibration weighting have stronger correlation to the response variable. (3) Both the first and the second order Rao-Scott corrections (entries under II and III) provide similar results compared to the ones using the actual limiting distributions (entries under I). (4) The validity of the replication weights is justified for cases with small sampling fractions but the results based on the estimated eigenvalues (entries under I, II and III) seem to work well even if $n/N=10\%$. (5) The bootstrap calibration method (entries under IV) works very well for $n/N=2\%$ for all cases. For cases with the large sampling fraction $n/N=10\%$, the size of the test for $H_0$: $\theta_{\None}=\theta_{\Ntwo}$ with $\sigma = \sigma_1$ is around $0.02$ for both PEL and SEL, showing the sensitivity of the replication weights on the bootstrap calibrated tests. (6) The Wald test has similar performance to SEL based tests in some cases but is less powerful in some other cases. 

Further investigation on the performance of the empirical likelihood methods for parameters defined through nonsmooth estimating functions is reported in the Appendix.

\medskip

\setcounter{section}{6}
\setcounter{equation}{0}
\no {\bf 6. An Application to the GSS 2016 Dataset}

\medskip

\no
The General Social Survey (GSS) is an annual cross-sectional survey conducted by Statistics Canada since 1985. The  survey gathers data on social trends in order to monitor changes in the living conditions and the well-being of Canadians, and to provide information on specific social policy issues. The 2016 GSS focused on Canadians at Work and Home, and collected information on the lifestyle behaviour of Canadians that affects their health and well-being, both in workplace and home. The survey covered individuals aged 15 years and older living in private households in the 10 provinces of Canada. Public-use GSS micro data files, which include the final survey weights and 500 sets of bootstrap weights, can be accessed through Statistics Canada's Research Data Centre (RDC) or the Data Liberation Initiative (DLI) at major Canadian universities. 

We analyzed a subset of the GSS 2016 data file using the pseudo empirical likelihood and the sample empirical likelihood methods developed in this paper. We explored the relationships between the response variable $y$ on job satisfaction and a set of 14 covariates through logistic regression analysis. The $y$ variable is dichotomized from the original 5-point likert scale, i.e., $y=1$ if either ``Very satisfied'' or ``Satisfied'' and $y=0$ otherwise. The set of covariates includes $x_1$: {\em Gender}; $x_2$: {\em Marital Status}; $x_3$: {\em Landed Immigrant Status}; $x_4$: {\em Citizenship Status}; $x_5$: {\em Number of Weeks Employed - Past 12 Months}; $x_6$: {\em Number of Weeks Worked at the Job - Past 12 Months}; $x_7$: {\em Unionized Job or Covered by Contract or Collective Agreement}; $x_8$: {\em Being Happy When Working Hard}; $x_9$: {\em Employment Benefits - Workplace Pension Plan}; $x_{10}$: {\em Employment Benefits - Paid Sick Leave}; $x_{11}$: {\em Employment Benefits - Paid Vacation Leave}; $x_{12}$: {\em Unfair Treatment/Discrimination - Past 12 Months}; $x_{13}$: {\em Age Group}; $x_{14}$: {\em Number of Persons Employed at Work Location}. The subset of the data file we used, denoted as $\mS$, consists of $n=1,552$ individuals who had valid responses to all 15 questions described above. Detailed descriptions of those questions are provided in the Supplementary Material. The final survey weights $w_i$ and the $b$th set of bootstrap weights $w_i^{(b)}$ are rescaled such that $\sum_{i\in \mS}w_i = n$ and $\sum_{i\in \mS}w_i^{(b)} = n$, $b=1,\cdots,500$. Note that the rescaling does not change the validity of the bootstrap weights for variance estimation as specified in Assumption 2. 

We considered the logistic regression model on $y$ given $x=(1,x_1,\cdots,x_{14})'$, which models ${\rm Pr}(y=1\mid x)$ through the logit link function  
${\rm logit} \{{\rm Pr}(y = 1\mid x)\}=x'\theta$, 
where ${\rm logit} (p)=\log\{p/(1-p)\}$ and $\theta=(\theta_0,\theta_1, \cdots,\theta_{14})'$. It follows that the odds for job satisfaction is given by 
$$
\frac{{\rm Pr}(y = 1\mid x)}{{\rm Pr}(y = 0\mid x)}=\prod_{j=0}^{14}\exp(x_j\theta_j).
$$
The value $\exp(\theta_j)$ represents the odds ratio (OR) for job satisfaction when $x_j$ changes from $0$ to $1$ given other covariates. 

The estimating function for defining $\theta$ is given by $g(x,y,\theta) = x\{y-\mu(x'\theta)\}$, where $\mu(x'\theta) = \exp(x'\theta)/\{1+\exp(x'\theta)\}$. \ Let \ $\theta_{\N} = (\theta_{\Nzero},\theta_{\None},\cdots,\theta_{\Nft})'$ \ be the finite population parameters defined by the census estimating equations. We computed the point estimates, the standard errors (SE), the odds ratios (OR) and the p-values for testing $H_0$: $\theta_{\Nj} = 0$ versus $H_1$: $\theta_{\Nj} \neq 0$, $j=0,1,\cdots,14$ using the pseudo empirical likelihood (PEL) and the sample empirical likelihood (SEL) methods. Note that we have $r=p$ in this case and the point estimates, the SE and the OR are the same under the two methods. The p-values for hypothesis tests were computed using the first-order Rao-Scott correction as described at the end of Section 3.2. The SCAD penalty function proposed by Fan and Li (2001) was used for variable selection. 

Results of estimation, hypothesis testing and variable selection are presented in Table \ref{tab-gss}. The first major observation is that the pseudo empirical likelihood and the sample empirical likelihood provide similar results for almost all cases, with only one noticeable exception on the p-value for testing $H_0$: $\theta_{\Ntw} = 0$. The second observation is that only three covariates, $x_8$: {\em Being Happy When Working Hard}, $x_{10}$:  {\em Employment Benefits - Paid Sick Leave}, and $x_{12}$: {\em Unfair Treatment/Discrimination - Past 12 Months}, show significance to the response variable on job satisfaction from individual tests given all other covariates in the model. The variable selection results, however, point to the fact that $x_8$ is the most significant factor on job satisfaction. 

\medskip

\setcounter{section}{7}
\setcounter{equation}{0}
\no {\bf 7. Additional Remarks}

\medskip

\no
Public-use survey data files might be utilized by researchers with diverse backgrounds and for different scientific objectives. Descriptive population parameters such as means and proportions, especially at the level of user-defined domains, are often of interest. However, complex survey data have also been used for analytic purposes. One important application is hypothesis tests in the presence of nuisance parameters. Binder and Patak (1994) discussed an estimating equation based test on one parameter in the presence of another nuisance parameter. Oguz-Alper and Berger (2016) presented a profile empirical likelihood test with nuisance parameters under the setting that detailed design information such as the first order inclusion probabilities and the population auxiliary information are available. They showed that the limiting distribution of the empirical likelihood ratio statistic follows a standard chisquare for certain sampling designs. General results, such as Theorems 1-6 presented in Section 3, for public-use survey data are not available in the existing literature. More importantly, naively assuming standard chisquare limiting distributions for the empirical likelihood ratio test statistics for public-use survey data files lead to invalid results as shown by the simulation  results presented in Table \ref{tab0}. 

A very important practical problem is variable selection when the survey dataset is used to fit a model involving a large number of covariates. The design-based variable selection techniques described in Section 3.4 are a major contribution of the current paper. Another topic of interest is to test the correctness of the specified model, which is equivalent to testing the unbiasedness of the estimating functions used in the constraints. A pseudo empirical likelihood or a sample empirical likelihood ratio test following Corollary 4 of Qin and Lawless (1994) seems to be possible. Detailed procedures are currently under investigation.

The empirical likelihood methods have been an active research topic during the past three decades, with many new developments covering different areas.  Rao and Wu (2009) contained an overview of empirical likelihood for complex surveys up to 2009. There have been several advances in recent years on empirical likelihood for complex surveys as evidenced by the additional  references cited in this paper. Reid (2012) provided an overview of likelihood inference in complex settings, and the development of empirical likelihood method for complex survey data received high attention on her list. Our paper addresses a topic with both theoretical and practical importance on analysis of public-use survey data files. Our proposed methods are valid for any public-use survey data files regardless of the original survey design. However, the bootstrap calibrated tests described in Section 4 put restrictions on how the final replication weights should be produced. Creating final replication weights for valid variance estimation (Assumption 2) has been known to be a challenging task at the data file production stage for complex surveys involving stratification and multi-stage unequal probability sampling. Our simulation results show that constructing replication weights to satisfy the requirements for the bootstrap calibration method is even harder. 
Another important topic is on how to handle item nonresponse for public-use data files. Single imputation methods are a popular approach among some statistical agencies to produce a single complete data file for public users. How to create replication weights for data files in the presence of imputation for missing values is a topic that deserves high attention in future research.

\medskip

\setcounter{equation}{0}
\def\theequation{A.\arabic{equation}}

\no {\bf Appendix}

\medskip

\no
\textbf{A.1 \ Lemmas}

\medskip

\no
We provide proofs of the main theoretical results presented in Section 3. 
To facilitate the development of large sample theories under the design-based framework, we rewrite the pseudo empirical log-likelihood function for a given  $\theta$ as 
\[
r_{\PEL}(\theta,\lambda) =   - n \sum_{i\in \mS} \tilde{w}_i(\mS) \log\bigl\{1+\lambda' g_i(\theta)\bigr\} \,,
\]
where the Lagrange multiplier $\lambda$ solves $g_{\PEL}(\lambda)=0$ given in equation (7) of the main paper. 
Let $\hat{\Lambda}_{\PEL}(\theta) = \{\lambda \mid \lambda'g_i(\theta)>-1, i\in \mS\}$. The range for $\lambda$ is defined by the constraints 
$\hat{p}_i(\theta) = \tilde{w}_i/\{1+\lambda'g_i(\theta)\}>0$ for all $i \in \mS$. 
The maximum pseudo empirical likelihood estimator of $\theta_{\N}$ is given by $
\hat\theta_{\PEL}=\arg\sup_{\theta\in\Theta}\min_{\lambda\in\hat{\Lambda}_{\PEL}(\theta)}r_{\PEL}(\theta,\lambda)$.
Let  $\hat\lambda_{\PEL}=\arg \min_{\lambda\in\hat{\Lambda}_n(\hat\theta_{\PEL})}r_{\SEL}(\hat\theta_{\PEL},\lambda)$.

Similar notation is introduced for the sample empirical likelihood function. 
Let $f_{\N}=n/N$ and $\hat{\Lambda}_{\SEL}(\theta) = \{\lambda \mid \lambda'w_if_{\N}g_i(\theta)>-1, i\in \mS\}$.  We rewrite the sample empirical log-likelihood function for a given  $\theta$ as
\[
r_{\SEL}(\theta,\lambda) = -\sum_{i\in \mS}\log\{1+\lambda'w_if_{\N}g_i(\theta)\} \,,
\]
where the Lagrange multiplier $\lambda$ solves $g_{\SEL}(\lambda)=0$ given in equation (9) of the main paper. 
The maximum sample empirical likelihood estimator is equivalently given by
$
\hat\theta_{\SEL}=\arg\sup_{\theta\in\Theta}\min_{\lambda\in\hat{\Lambda}_{\SEL}(\theta)}r_{\SEL}(\theta,\lambda)$.
Let  $\hat\lambda_{\SEL}=\arg \min_{\lambda\in\hat{\Lambda}_n(\hat\theta_{\SEL})}r_{\SEL}(\hat\theta_{\SEL},\lambda)$. Let ``w.p.a.1'' denote ``with probability approaching 1''. 

The following three lemmas  are required for establishing the asymptotic normality of our proposed maximum pseudo and sample empirical likelihood estimators. Proofs of the lemmas follow similar arguments used in Zhao, Haziza and Wu (2018). Details are omitted.

\medskip

\noindent
{\bf Lemma 1.} {\em 
Suppose that Assumptions  1, 3, 4 and 5 hold. Let $\Lambda_n = \{\lambda \mid \|\lambda\| \leq cn^{-1/2}\}$ for a given $c>0$. Then \\
(i)
$\sup_{\theta \in \Theta, \lambda \in \Lambda_n, i \in \mS}|\lambda'
g_i(\theta)| =o_p(1)$, and  with probability approaching 1, $\Lambda_n\subseteq\hat{\Lambda}_{\PEL}(\theta)$  for all $\theta\in\Theta$;  \\
(ii)
$\sup_{\theta \in \Theta, \lambda \in \Lambda_n, i \in \mS}|\lambda'w_if_{\N}
g_i(\theta)| =o_p(1)$, and  with probability approaching 1, $\Lambda_n\subseteq\hat{\Lambda}_{\SEL}(\theta)$  for all $\theta\in\Theta$.
}

\medskip

\noindent
{\bf Lemma 2.} {\em 
Suppose that Assumptions 1, 3, 4 and 5 hold and that $\bar{\theta}\in\Theta$, $\bar{\theta}\stackrel{p}{\rightarrow} \theta_{\N}$ and $\|\hat{U}_{n}(\bar{\theta})\| = O_p(n^{-1/2})$. Then, with $h$ indicating either $PEL$ or $SEL$, $\bar\lambda=\arg \sup_{\lambda\in\hat{\Lambda}_{h}(\bar\theta)}r_h(\bar\theta,\lambda)$ exists w.p.a.1, $\bar\lambda=O_p(n^{-1/2})$, and $\sup_{\lambda\in\hat{\Lambda}_h(\bar\theta)}r_h(\bar\theta,\lambda)\leq O_p(n^{-1})$.
}

\medskip

\noindent
{\bf Lemma 3.} {\em 
Suppose that Assumptions  1, 3, 4 and 5 hold. Then, with $h$ indicating either $PEL$ or $SEL$,   $\|\hat{U}_{n}(\hat\theta_{h})\| = O_p(n^{-1/2})$ and $\|\hat{U}_{n}(\hat\theta_{h})\| = O_p(n^{-1/2})$.
}

\medskip

\no 
{\bf   A.2 \ Proof of Theorem 1}

\medskip

\no
The proof has similarities to the proof of Theorem 1 in Qin and Lawless (1994). Define
\[
\begin{array}{lll}
Q_{n1}(\theta,\lambda)&=&\sum\limits_{i\in \mS}\frac{\tilde{w}_i(\mS)g_i(\theta)}
{1+\lambda'g_i(\theta)},\\
Q_{n2}(\theta,\lambda)&=&\sum\limits_{i\in \mS}\frac{\tilde{w}_i(\mS)}
{1+\lambda'g_i(\theta)}\Big\{\frac{\partial g_i(\theta)}{\partial\theta'}\Big\}'\lambda \,.
\end{array}
\]
Then $\hat{\theta}_{\PEL}$ and $\hat{\lambda}_{\PEL}$ satisfy
$$
Q_{n1}(\hat{\theta}_{\PEL},\hat{\lambda}_{\PEL})=0, \;\;\;\;\; Q_{n2}(\hat{\theta}_{\PEL},\hat{\lambda}_{\PEL})=0 \,.
$$
Taking the Taylor expansion of $Q_{n1}(\hat{\theta}_{\PEL},\hat{\lambda}_{\PEL})$ and $Q_{n2}(\hat{\theta}_{\PEL},\hat{\lambda}_{\PEL})$ around  $(\theta_{\N},0)$ yields
\[\begin{array}{lll}
0&=&Q_{n1}(\hat{\theta}_{\PEL},\hat{\lambda}_{\PEL})\\
&=&Q_{n1}(\theta_{\N},0)+\frac{\partial Q_{n1}(\theta_{\N},0)}{\partial \theta'}(\hat{\theta}_{\PEL}-\theta_{\N})+\frac{\partial Q_{n1}(\theta_{\N},0)}{\partial \lambda'}(\hat{\lambda}_{\PEL}-0)+o_p(\sigma_n),\\
0&=&Q_{n2}(\hat{\theta}_{\PEL},\hat{\lambda}_{\PEL})\\
&=&Q_{n2}(\theta_{\N},0)+\frac{\partial Q_{n2}(\theta_{\N},0)}{\partial \theta'}(\hat{\theta}_{\PEL}-\theta_{\N})+\frac{\partial Q_{n2}(\theta_{\N},0)}{\partial \lambda'}(\hat{\lambda}_{\PEL}-0)+o_p(\sigma_n),
\end{array}
\]
where $\sigma_n=\| \hat{\theta}_{\PEL}-\theta_{\N}\|+\| \hat{\lambda}_{\PEL}\|$.
It can be shown that the four terms involved in the above equations are given by 
\begin{eqnarray*}
\frac{\partial Q_{n1}(\theta_{\N},0)}{\partial \theta'}& = &\sum\limits_{i\in \mS}\tilde{w}_i(\mS)\frac{\partial g_i(\theta)}{\partial\theta'}\Big|_{\theta=\theta_{\N}} \,, \\
\frac{\partial Q_{n1}(\theta_{\N},0)}{\partial \lambda'}&=&-\sum\limits_{i\in \mS}\tilde{w}_i(\mS)g_i(\theta_{\N})g_i(\theta_{\N})' \,, \\
\frac{\partial Q_{n2}(\theta_{\N},0)}{\partial \theta'}& = &0 \,, \\
\frac{\partial Q_{n2}(\theta_{\N},0)}{\partial \lambda'}&=&\sum\limits_{i\in \mS}\tilde{w}_i(\mS)\Big\{\frac{\partial g_i(\theta)}{\partial\theta'}\Big\}'\Big|_{\theta=\theta_{\N}} \,.
\end{eqnarray*}
We have
\[\left(\begin{array}{c}
\hat{\lambda}_{\PEL}\\
\hat{\theta}_{\PEL}-\theta_{\N}
\end{array}\right)=
S_{n1}^{-1}\left(\begin{array}{c}
-Q_{n1}(\theta_{\N},0)+o_p(\sigma_n)\\
o_p(\sigma_n)
\end{array}\right),
\]
where
\[\begin{array}{lllll}
S_{n1}&=&\left(\begin{array}{cc}
-\sum\limits_{i\in \mS}\tilde{w}_i(\mS)g_i(\theta_{\N})g_i(\theta_{\N})'&\sum\limits_{i\in \mS}\tilde{w}_i(\mS)\frac{\partial g_i(\theta)}{\partial\theta'}\Big|_{\theta=\theta_{\N}}\\
\sum\limits_{i\in \mS}\tilde{w}_i(\mS)\Big\{\frac{\partial g_i(\theta)}{\partial\theta'}\Big\}'\Big|_{\theta=\theta_{\N}}&0
\end{array}\right)
\\
\stackrel{p}{\rightarrow}S &=&
\left(\begin{array}{ccccc}
-W_1&&&&\Gamma\\
\Gamma'&&&&0
\end{array}\right).
\end{array}
\]

Noting that $Q_{n1}(\theta_{\N},0)=\sum_{i\in \mS}\tilde{w}_i(\mS)g_i(\theta_{\N})=O_p(n^{-1/2})$, it can be shown that $\sigma_n=O_p(n^{-1/2})$. It follows  that
\[\begin{array}{lllll}
\hat\theta_{\PEL}-\theta_{\N}&=&-\Sigma_1\Gamma'W_1^{-1}Q_{n1}(\theta_{\N},0)+o_p(n^{-1/2}),\\
\hat\lambda_{\PEL} &=&P_1Q_{n1}(\theta_{\N},0)+o_p(n^{-1/2}),\\
\end{array}
\]
where $\Sigma_1=(\Gamma' W_1^{-1}\Gamma)^{-1}$ and $P_1=W_1^{-1}-W_1^{-1}\Gamma\Sigma_1\Gamma'W_1^{-1}$.
Combining above arguments  with Assumption 1, the asymptotic normality of the estimator $\hat\theta_{\PEL}$ is established. This completes the proof of Theorem 1.

\medskip

\no 
{\bf   A.3 \ Proof of Theorem 2}

\medskip

\no
Denote $\lambda_{\N}=\lambda(\theta_{\N})$, which is  the solution to
\[
g_{\PEL} (\lambda_{\N}) = \sum_{i\in {\mS}} \frac{\tilde{w}_i({\mS}) g_i(\theta_{\N})}{1+\lambda_{\N}' g_i(\theta_{\N})} = 0 \,.
\label{lam1}
\]
Applying the Taylor series expansion to $g_{\PEL} (\lambda_{\N})$ around $\lambda_{\N}=0$, together with Lemmas 1-3, we have that
\[\begin{array}{lllll}
\lambda_{\N} &=&\Bigl[\sum\limits_{i\in \mS} \tilde{w}_i(\mS) g_i(\theta_{\N})g_i(\theta_{\N})' \Bigr]^{-1} \sum\limits_{i\in \mS} \tilde{w}_i(\mS) g_i(\theta_{\N}) + o_p\Bigl( n^{-1/2}\Bigr)\\
&=&W_1^{-1} (1/N) \sum\limits_{i\in \mS} w_i g_i(\theta_{\N}) + o_p\Bigl( n^{-1/2}\Bigr) \,.
\end{array}
\]
This leads to the following asymptotic expansion to the pseudo empirical log-likelihood ratio statistic:
\[\begin{array}{lllll}
-2r_{\PEL}(\theta_{\N},\lambda_{\N})& =& n \lambda_{\N}'W_1\lambda_{\N} + o_p\bigl(1\bigr)\\
&=&(n/N^2)\hat{U}_n(\theta_{\N})'W_1^{-1}\hat{U}_n(\theta_{\N})+ o_p\bigl(1\bigr).
\end{array}\]
where $
\hat{U}_n(\theta) = \sum_{i\in \mS} w_i \, g(x_i,y_i,\theta).
$ Note that $P_1W_1P_1=P_1$. This, coupled with the proof of Theorem 1, shows that
\[\begin{array}{lllll}
-2r_{\PEL}(\hat\theta_{\PEL},\hat\lambda_{\PEL})& =&
 = n\hat{\lambda}_{\PEL}'W_1\hat{\lambda}_{\PEL}+o_p(1)\\
&=&(n/N^2)\hat{U}_n(\theta_{\N})'P_1W_1 P_1\hat{U}_n(\theta_{\N})+o_p(1)\\
&=&(n/N^2)\hat{U}_n(\theta_{\N})'P_1\hat{U}_n(\theta_{\N})+o_p(1).
\end{array}
\]
By Assumption 1, it can be shown that
$(\sqrt{n}/N)\hat{U}_n(\theta_{\N})$
is asymptotically normally distributed with mean zero  and  variance-covariance matrix at the order $O(1)$.
Combining above arguments, we can show that
\[\begin{array}{lllll}
{\rm LR}_{\PEL}(\theta_{\N}) &=& 2\big\{r_{\PEL}(\hat\theta_{\PEL},\hat\lambda_{\PEL})-r_{\PEL}(\theta_{\N},\lambda_{\N})\big\}\\
&=&(n/N^2)\hat{U}_n(\theta_{\N})'W_1^{-1}\Gamma\Sigma_1  \Gamma'W_1^{-1}\hat{U}_n(\theta_{\N})+o_p(1)\\
&\stackrel{{\cal L}}{\rightarrow}&Q'\Omega^{1/2}W_1^{-1}\Gamma\Sigma_1\Gamma'W_1^{-1}\Omega^{1/2}Q,
\end{array}
\]
where $Q \sim N(0, I_r)$,
$I_r$ is the $r\times r$ identity matrix, $r$ is the dimension of population estimating equations.
This completes the proof of Theorem 2.

\medskip

\no 
{\bf   A.4 \ Proof of Theorem 3}

\medskip

\no
Define $\Phi(\theta)=\partial R(\theta)/\partial \theta'$ which is a $k\times p$ matrix. We first derive the asymptotic distribution of $\hat\theta_{\PEL}^{\R}$.
Note that finding the maximizer
$\hat\theta_{\PEL}^{\R}=\mathop{\arg\max}_{\theta\in \Theta^{\R}}r_{\PEL}(\theta)$ is equivalent to optimizing  the following objective function
\[
\begin{array}{lll}
r_{\PEL}^{\R}(\theta,\lambda,\tau)=
\sum\limits_{i\in \mS}\tilde{w}_i(\mS)\log\{1+\lambda'g_i(\theta)\}
+\tau'R(\theta)
\end{array}
\]
with respect to $(\theta,\lambda,\tau)$, where  $\tau$ is another $k\times 1$ vector of Lagrange multiplier for the constrained maximization. The optimizer $(\hat\theta_{\PEL}^{\R},\hat{\lambda}_{\PEL}^{\R},\hat{\tau}_{\PEL}^{\R})$ of $r_{\PEL}^{\R}(\theta,\lambda,\tau)$ satisfies $0=Q_{nj}^{\R}(\hat\theta_{\PEL}^{\R},\hat{\lambda}_{\PEL}^{\R},\hat{\tau}_{\PEL}^{\R})$ for $j=1,2,3$, where
\[
\begin{array}{lll}
Q_{n1}^{\R}(\theta,\lambda,\tau)&=&\sum\limits_{i\in \mS}\frac{\tilde{w}_i(\mS)g_i(\theta)}
{1+\lambda'g_i(\theta)},\\
Q_{n2}^{\R}(\theta,\lambda,\tau)&=&\sum\limits_{i\in \mS}\frac{\tilde{w}_i(\mS)}
{1+\lambda'g_i(\theta)}\Big\{\frac{\partial g_i(\theta)}{\partial\theta'}\Big\}'\lambda+\Phi(\theta)'\tau,\\
Q_{n3}^{\R}(\theta,\lambda,\tau)&=&R(\theta).
\end{array}
\]
It can be shown through direct calculations that
\begin{eqnarray*}
\frac{\partial Q_{n1}^{\R}(\theta_{\N},0,0)}{\partial \theta'}
&=&\sum\limits_{i\in \mS}\tilde{w}_i(\mS)\frac{\partial g_i(\theta)}{\partial\theta}\Big|_{\theta=\theta_{\N}}, \\
\frac{\partial Q_{n1}^{\R}(\theta_{\N},0,0)}{\partial \lambda'}
&=&-\sum\limits_{i\in \mS}\tilde{w}_i(\mS)g_i(\theta_{\N})g_i(\theta_{\N})',\\
\frac{\partial Q_{n1}^{\R}(\theta_{\N},0,0)}{\partial \tau'}&=&0,\\
\frac{\partial Q_{n2}^{\R}(\theta_{\N},0,0)}{\partial \theta'}&=&0,\\
\frac{\partial Q_{n2}^{\R}(\theta_{\N},0,0)}{\partial \lambda'}
&=&\sum\limits_{i\in \mS}\tilde{w}_i(\mS)\Big\{\frac{\partial g_i(\theta)}{\partial\theta'}\Big\}'\Big|_{\theta=\theta_{\N}},\\
\frac{\partial Q_{n2}^{\R}(\theta_{\N},0,0)}{\partial \tau'}&=&\Phi(\theta_{\N})',\\
\frac{\partial Q_{n3}^{\R}(\theta_{\N},0,0)}{\partial \theta'}&=&\Phi(\theta_{\N}),\\
\frac{\partial Q_{n3}^{\R}(\theta_{\N},0,0)}{\partial \lambda'}&=&0,\\
\frac{\partial Q_{n3}^{\R}(\theta_{\N},0,0)}{\partial \tau'}&=&0.
\end{eqnarray*}
Using a multivariate Taylor series expansion to $Q_{nj}(\hat\theta_{\PEL}^{\R},\hat{\lambda}_{\PEL}^{\R},\hat{\tau}_{\PEL}^{\R})$ at $(\theta_{\N},0,0)$, we have 
$$
\left(\begin{array}{ccccc}
-W_1&\Gamma&0\\\Gamma'&0&\Phi'\\
0&\Phi&0\end{array}\right)
\left(\begin{array}{ccccc}\hat{\lambda}_{\PEL}^{\R}\\
\hat\theta_{\PEL}^{\R} - \theta_{\N}\\\hat{\tau}_{\PEL}^{\R}\end{array}\right) = \left(\begin{array}{ccccc}
-Q_{n1}^{\R}(\theta_{\N},0,0)\\
0\\0\end{array}\right)+o_p(n^{-1/2}),
$$
where  $\Phi=\Phi(\theta_{\N})$. Now define
$$
H=\left(\begin{array}{ccccc}
-W_1&\Gamma&0\\\Gamma'&0&\Phi'\\
0&\Phi&0\end{array}\right)=:\left(\begin{array}{ccccc}
H_{11}&H_{12}\\H_{21}&H_{22}\end{array}\right) \,,
$$
where $H_{11}=-W_1$, $H_{12}=(\Gamma,0)$, $H_{21}=H_{12}'$ and
$$
H_{22}=\left(\begin{array}{ccccc}
0&\Phi'\\
\Phi&0\end{array}\right) \,.
$$
Applying the theory of  block matrix inversions, we obtain
$$
H^{-1}=\left(\begin{array}{ccccc}
H_{11}^{-1}&0\\
0&0\end{array}\right)+\left(\begin{array}{ccccc}
-H_{11}^{-1}H_{12}\\
I\end{array}\right)K^{-1}(-H_{21}H_{11}^{-1}~~ I) \,,
$$
where
$$
K=H_{22}-H_{21}H_{11}^{-1}H_{12}=\left(\begin{array}{ccccc}
\Sigma_1^{-1} &\Phi'\\
\Phi&0\end{array}\right) \,.
$$
In addition,  we also have that
$$
K^{-1}=\left(\begin{array}{ccccc}
\Sigma_1-\Sigma_1 \Phi'(\Phi\Sigma_1 \Phi')^{-1}\Phi\Sigma_1&-\Sigma_1 \Phi'(\Phi\Sigma_1 \Phi')^{-1}\\
-(\Phi\Sigma_1 \Phi')^{-1}\Phi\Sigma_1 &(\Phi\Sigma_1 \Phi')^{-1}\end{array}\right)\,.
$$
This leads to 
$$
\left(\begin{array}{ccccc}
\hat{\theta}_{\PEL}^{\R} - \theta_{\N}\\\hat{\tau}_{\PEL}^{\R}\end{array}\right) =K^{-1}H_{21}H_{11}^{-1}Q_{n1}^{\R}(\theta_{\N},0,0)+ o_p(n^{-1/2}) \,,
$$
and 
$$
\hat{\lambda}_{\PEL}^{\R}=-[H_{11}^{-1}+H_{11}^{-1}H_{12}K^{-1}H_{21}H_{11}^{-1}]Q_{n1}^{\R}(\theta_{\N},0,0)+ o_p(n^{-1/2}) \,.
$$
It further leads to 
\begin{eqnarray*}
\hat{\theta}_{\PEL}^{\R}-\theta_{\N} &=& -P_1^{\R}\Gamma'W_1^{-1}Q_{n1}^{\R}(\theta_{\N},0,0)+o_p(n^{-1/2}) \,, \\
\hat{\lambda}_{\PEL}^{\R}  &=& P_2^{\R}Q_{n1}^{\R}(\theta_{\N},0,0)+o_p(n^{-1/2}) \,, \\
\hat{\tau}_{\PEL}^{\R} &=& (\Phi\Sigma_1 \Phi')^{-1}\Phi\Sigma_1\Gamma'W_1^{-1}Q_{n1}^{\R}(\theta_{\N},0,0)+o_p(n^{-1/2}) \,,
\end{eqnarray*}
where $P_1^{\R}=\Sigma_1-\Sigma_1 \Phi'(\Phi\Sigma_1 \Phi')^{-1}\Phi\Sigma_1$ and $P_2^{\R} = W_1^{-1}-W_1^{-1}\Gamma P_1^{\R}\Gamma'W_1^{-1}$.

We now derive the asymptotic distribution of the empirical log-likelihood ratio statistic
\[
{\rm LR}_{\PEL}^{\R}(\hat\theta_{\PEL}^{\R}) = 2\big\{r_{\PEL}(\hat\theta_{\PEL},\hat\lambda_{\PEL})-r_{\PEL}(\hat\theta_{\PEL}^{\R},\hat\lambda_{\PEL}^{\R})\big\} \,.
\] 
Noting that $P_2^{\R}W_1 P_2^{\R}=P_2^{\R}$, we have 
\[\begin{array}{lllll}
&& -2r_{\PEL}(\hat\theta_{\PEL}^{\R},\hat\lambda_{\PEL}^{\R}) \\
& =& 2 n \sum_{i\in \mS} \tilde{w}_i(\mS) \log\bigl\{1+\hat{\lambda}_{\PEL}^{\R} g_i(\hat\theta_{\PEL}^{\R})\bigr\}
 = n\hat{\lambda}_{\PEL}^{*\top}W_1\hat{\lambda}_{\PEL}^{\R}+o_p(1)\\
&=&nQ_{n1}^{*\top}(\theta_{\N},0,0)P_2^{\R}W_1 P_2^{\R}Q_{n1}^{\R}(\theta_{\N},0,0)+o_p(1)\\
&=&nQ_{n1}^{*\top}(\theta_{\N},0,0)P_2^{\R}Q_{n1}^{\R}(\theta_{\N},0,0)+o_p(1).
\end{array}
\]
From the proof of Theorem 1, we have
$$
-2r_{\PEL}(\hat\theta_{\PEL},\hat\theta_{\SEL})
=nQ_{n1}'(\theta_{\N},0)P_1Q_{n1}(\theta_{\N},0)+o_p(1).
$$
Then,
\[\begin{array}{lllll}
&& {\rm LR}_{\PEL}^{\R}(\hat\theta_{\PEL}^{\R})  \\
&=& 2\big\{r_{\PEL}(\hat\theta_{\PEL},\hat\lambda_{\PEL})
-r_{\PEL}(\hat\theta_{\PEL}^{\R},\hat\lambda_{\PEL}^{\R})\big\} \\
&=& nQ_{n1}'(\theta_{\N},0)(P_2^{\R}-P_1)Q_{n1}(\theta_{\N},0) + o_p(1) \,\\
&=&nQ_{n1}'(\theta_{\N},0)W_1^{-1}\Gamma(\Sigma_1-P_1^{\R})\Gamma'W_1^{-1}Q_{n1}(\theta_{\N},0)+o_p(1)\\
&=&nQ_{n1}'(\theta_{\N},0)W_1^{-1}\Gamma\Sigma_1 \Phi'(\Phi\Sigma_1 \Phi')^{-1}\Phi\Sigma_1\Gamma'W_1^{-1}Q_{n1}(\theta_{\N},0)
+o_p(1).
\end{array}
\]
By Assumption 1, it can be shown that
$$
n^{1/2}Q_{n1}(\theta_{\N},0)=n^{1/2}\sum_{i\in \mS}\tilde{w}_i(\mS)g_i(\theta_{\N})\stackrel{{\cal L}}{\rightarrow}N(0,\Omega) \,,
$$
where $\Omega=(n/N^2)Var\{\sum_{i\in \mS}w_ig_i(\theta_{\N})\mid {\mF}_{\N}\}$.
Therefore,
\[\begin{array}{lllll}
{\rm LR}_{\PEL}^{\R}(\hat\theta_{\PEL}^{\R})\mid {\mF}_{\N} \stackrel{{\cal L}}{\rightarrow}Q'\Delta_1^{\R}Q \,,
\end{array}
\]
where $Q \sim N(0, I_r)$ and
$
\Delta_1^{\R} = \Omega^{1/2}W_1^{-1}\Gamma\Sigma_1 \Phi'(\Phi\Sigma_1 \Phi')^{-1}\Phi\Sigma_1\Gamma'W_1^{-1}\Omega^{1/2} \,
$
with
$\Sigma_1=(\Gamma' W_1^{-1}\Gamma)^{-1}$. The proof of Theorem 3 is then completed.

\medskip

\no 
{\bf   A.5 \ Proof of Theorem 4}

\medskip

\no
Major steps of the proof are similar to the proof of Theorem 1. If we define
\[
\begin{array}{lll}
M_{n1}(\theta,\lambda)&=&\frac{1}{n}\sum\limits_{i\in \mS}\frac{w_if_{\N}g_i(\theta)}
{1+\lambda'w_if_{\N}g_i(\theta)},\\
M_{n2}(\theta,\lambda)&=&\frac{1}{n}\sum\limits_{i\in \mS}\frac{w_if_{\N}}
{1+\lambda'w_if_{\N}g_i(\theta)}\Big\{\frac{\partial g_i(\theta)}{\partial\theta'}\Big\}'\lambda \,,
\end{array}
\]
then $\hat{\theta}_{\SEL}$ and $\hat{\lambda}_{\SEL}$ satisfy
$$
M_{n1}(\hat{\theta}_{\SEL},\hat{\lambda}_{\SEL})=0, \;\;\;\;\; M_{n2}(\hat{\theta}_{\SEL},\hat{\lambda}_{\SEL})=0 \,.
$$

Taking the Taylor series expansion of $M_{n1}(\hat{\theta}_{\SEL},\hat{\lambda}_{\SEL})$ and $M_{n2}(\hat{\theta}_{\SEL},\hat{\lambda}_{\SEL})$ at $(\theta_{\N},0)$ yields
\[\begin{array}{lll}
0&=&M_{n1}(\hat{\theta}_{\SEL},\hat{\lambda}_{\SEL})\\
&=&M_{n1}(\theta_{\N},0)+\frac{\partial M_{n1}(\theta_{\N},0)}{\partial \theta'}(\hat{\theta}_{\SEL}-\theta_{\N})+\frac{\partial M_{n1}(\theta_{\N},0)}{\partial \lambda'}(\hat{\lambda}_{\SEL}-0)+o_p(\sigma_n),\\
0&=&M_{n2}(\hat{\theta}_{\SEL},\hat{\lambda}_{\SEL})\\
&=&M_{n2}(\theta_{\N},0)+\frac{\partial M_{n2}(\theta_{\N},0)}{\partial \theta'}(\hat{\theta}_{\SEL}-\theta_{\N})+\frac{\partial M_{n2}(\theta_{\N},0)}{\partial \lambda'}(\hat{\lambda}_{\SEL}-0)+o_p(\sigma_n),
\end{array}
\]
where $\sigma_n=\| \hat{\theta}_{\SEL}-\theta_{\N}\|+\| \hat{\lambda}_{\SEL}\|$.
By direct calculation, we obtain
\begin{eqnarray*}
\frac{\partial M_{n1}(\theta_{\N},0)}{\partial \theta'}
& = &\frac{1}{n}\sum\limits_{i\in \mS}w_if_{\N}\frac{\partial g_i(\theta)}{\partial\theta'}\Big|_{\theta=\theta_{\N}},\\
\frac{\partial M_{n1}(\theta_{\N},0)}{\partial \lambda'}&=&-\frac{1}{n}\sum\limits_{i\in \mS}w_i^2f_{\N}^2g_i(\theta_{\N})g_i(\theta_{\N})' \,, \\
\frac{\partial M_{n2}(\theta_{\N},0)}{\partial \theta'}& = &0 \,, \\
\frac{\partial M_{n2}(\theta_{\N},0)}{\partial \lambda'}&=&\frac{1}{n}\sum\limits_{i\in \mS}w_if_{\N}\Big\{\frac{\partial g_i(\theta)}{\partial\theta'}\Big\}'\Big|_{\theta=\theta_{\N}} \,.
\end{eqnarray*}
This leads to 
\[\left(\begin{array}{c}
\hat{\lambda}_{\SEL}\\
\hat{\theta}_{\SEL}-\theta_{\N}
\end{array}\right)=
S_{n2}^{-1}\left(\begin{array}{c}
-M_{n1}(\theta_{\N},0)+o_p(\sigma_n)\\
o_p(\sigma_n)
\end{array}\right),
\]
where
\[\begin{array}{lllll}
S_{n2}&=&\left(\begin{array}{cc}
-\frac{1}{n}\sum\limits_{i\in \mS}w_i^2f_{\N}^2g_i(\theta_{\N})g_i(\theta_{\N})'&\frac{1}{n}\sum\limits_{i\in \mS}w_if_{\N}\frac{\partial g_i(\theta)}{\partial\theta'}\Big|_{\theta=\theta_{\N}}\\
\frac{1}{n}\sum\limits_{i\in \mS}w_if_{\N}\Big\{\frac{\partial g_i(\theta)}{\partial\theta'}\Big\}'\Big|_{\theta=\theta_{\N}}&0
\end{array}\right)
\\
\stackrel{p}{\rightarrow}S_2 &=&
\left(\begin{array}{ccccc}
-W_2&&&&\Gamma\\
\Gamma'&&&&0
\end{array}\right).
\end{array}
\]
Noting that $M_{n1}(\theta_{\N},0)=(1/n)\sum_{i\in \mS}w_if_{\N}g_i(\theta_{\N})=O_p(n^{-1/2})$, it can be shown that $\sigma_n=O_p(n^{-1/2})$. 
We have 
\[\begin{array}{lllll}
\hat\theta_{\SEL}-\theta_{\N}&=&-\Sigma_2\Gamma'W_2^{-1}M_{n1}(\theta_{\N},0)+o_p(n^{-1/2}),\\
\hat\lambda_{\SEL} &=&P_2M_{n1}(\theta_{\N},0)+o_p(n^{-1/2}),\\
\end{array}
\]
where $\Sigma_2=(\Gamma' W_2^{-1}\Gamma)^{-1}$ and $P_2=W_2^{-1}-W_2^{-1}\Gamma\Sigma_2\Gamma'W_2^{-1}$.
The proof of Theorem 4 is then completed by combining above arguments  with Assumption 1.

\medskip

\no 
{\bf   A.6 \ Proof of Theorem 5}

\medskip

\no
The proof is similar to the proof of Theorem 2. Let   $\lambda_{\N}=\lambda(\theta_{\N})$ be  the solution to
\[
g_{\SEL}(\lambda_{\N}) = \frac{1}{n}\sum\limits_{i\in \mS}\frac{w_if_{\N}g_i(\theta_{\N})}
{1+\lambda_{\N}'w_if_{\N}g_i(\theta_{\N})} = 0.
\label{lam1}
\]
Applying the Taylor series expansion to $g_{\SEL}(\lambda_{\N})$ around $\lambda_{\N}=0$, together with Lemmas 1-3, we have that
\[\begin{array}{lllll}
\lambda_{\N} &=&\Bigl[(n/N^2)\sum\limits_{i\in \mS} w_i^2 g_i(\theta_{\N})g_i(\theta_{\N})' \Bigr]^{-1} (1/N) \sum\limits_{i\in \mS} w_i g_i(\theta_{\N}) + o_p\Bigl( n^{-1/2}\Bigr)\\
&=&W_2^{-1} (1/N) \sum\limits_{i\in \mS} w_i g_i(\theta_{\N}) + o_p\Bigl( n^{-1/2}\Bigr) \,.
\end{array}
\]
By the Taylor series expansion of $-2nr_{\SEL}(\theta_{\N},\lambda_{\N})$ around $\lambda_{\N}=0$, we have 
\[\begin{array}{lllll}
-2nr_{\PEL}(\theta_{\N},\lambda_{\N}) = n \lambda_{\N}'W_2\lambda_{\N} + o_p\bigl(1\bigr)
=(n/N^2)\hat{U}_n(\theta_{\N})'W_2^{-1}\hat{U}_n(\theta_{\N})+ o_p\bigl(1\bigr).
\end{array}\]
It follows from  the proof of Theorem 4 that
\[\begin{array}{lllll}
-2nr_{\SEL}(\hat\theta_{\SEL},\hat\lambda_{\SEL})& =&
  n\hat{\lambda}_{\SEL}'W_2\hat{\lambda}_{\SEL}+o_p(1)\\
&=&(n/N^2)\hat{U}_n(\theta_{\N})'P_2W_2 P_2\hat{U}_n(\theta_{\N})+o_p(1)\\
&=&(n/N^2)\hat{U}_n(\theta_{\N})'P_2\hat{U}_n(\theta_{\N})+o_p(1).
\end{array}
\]
The last equality holds since $P_2W_2P_2=P_2$.
Combining above arguments, we can show that
\[\begin{array}{lllll}
{\rm LR}_{\SEL}(\theta_{\N}) &=& 2n\big\{r_{\SEL}(\hat\theta_{\SEL},\hat\lambda_{\SEL})-r_{\SEL}(\theta_{\N},\lambda_{\N})\big\}\\
&=&(n/N^2)\hat{U}_n(\theta_{\N})'W_2^{-1}\Gamma\Sigma_2  \Gamma'W_2^{-1}\hat{U}_n(\theta_{\N})+o_p(1)\\
&\stackrel{{\cal L}}{\rightarrow}&Q'\Omega^{1/2}W_2^{-1}\Gamma\Sigma_2\Gamma'W_2^{-1}\Omega^{1/2}Q.
\end{array}
\]
The proof of Theorem 5  is then completed.

\medskip

\no 
{\bf   A.7 \ Proof of Theorem 6}

\medskip

\no
The proof is similar to the proof of Theorem 3. We first derive the asymptotic distribution of $\hat\theta_{\SEL}^{\R}$.
Finding the maximizer
$\hat\theta_{\SEL}^{\R}=\mathop{\arg\max}_{\theta\in \Theta^{\R}}r_{\SEL}(\theta)$ is equivalent to optimizing  the following objective function
\[
\begin{array}{lll}
r_{\SEL}^{\R}(\theta,\lambda,\tau)=\frac{1}{n}
\sum\limits_{i\in \mS}\log\{1+\lambda'w_if_{\N}g_i(\theta)\}
+\tau'R(\theta)
\end{array}
\]
with respect to $(\theta,\lambda,\tau)$, where  $\tau$ is another $k\times 1$ vector of Lagrange multiplier for the constrained maximization. The optimizer $(\hat\theta_{\SEL}^{\R},\hat{\lambda}_{\SEL}^{\R},\hat{\tau}_{\SEL}^{\R})$ of $r_{\SEL}^{\R}(\theta,\lambda,\tau)$ satisfies 
\[
0=M_{nj}^{\R}(\hat\theta_{\SEL}^{\R},\hat{\lambda}_{\SEL}^{\R},\hat{\tau}_{\SEL}^{\R})\,, \;\;\;\; j=1,2,3 \,, 
\]
where
\[
\begin{array}{lll}
M_{n1}^{\R}(\theta,\lambda,\tau)&=&\frac{1}{n}\sum\limits_{i\in \mS}\frac{w_if_{\N}g_i(\theta)}
{1+\lambda'w_if_{\N}g_i(\theta)},\\
M_{n2}^{\R}(\theta,\lambda,\tau)&=&\frac{1}{n}\sum\limits_{i\in \mS}\frac{w_if_{\N}}
{1+\lambda'w_if_{\N}g_i(\theta)}\Big\{\frac{\partial g_i(\theta)}{\partial\theta'}\Big\}'\lambda+\Phi(\theta)'\tau,\\
M_{n3}^{\R}(\theta,\lambda,\tau)&=&R(\theta).
\end{array}
\]
It can be shown through direct calculations that
\begin{eqnarray*}
\frac{\partial M_{n1}^{\R}(\theta_{\N},0,0)}{\partial \theta'}&=&\frac{1}{n}\sum\limits_{i\in \mS}w_if_{\N}\frac{\partial g_i(\theta)}{\partial\theta'}\Big|_{\theta=\theta_{\N}},\\
\frac{\partial M_{n1}^{\R}(\theta_{\N},0,0)}{\partial \lambda'}&=&-\frac{1}{n}\sum\limits_{i\in \mS}w_i^2f_{\N}^2g_i(\theta_{\N})g_i(\theta_{\N})',\\
\frac{\partial M_{n1}^{\R}(\theta_{\N},0,0)}{\partial \tau}&=&0,\\
\frac{\partial M_{n2}^{\R}(\theta_{\N},0,0)}{\partial \theta'}&=&0,\\
\frac{\partial M_{n2}^{\R}(\theta_{\N},0,0)}{\partial \lambda'}&=&\frac{1}{n}\sum\limits_{i\in \mS}w_if_{\N}\Big\{\frac{\partial g_i(\theta)}{\partial\theta'}\Big\}'\Big|_{\theta=\theta_{\N}},\\
\frac{\partial M_{n2}^{\R}(\theta_{\N},0,0)}{\partial \tau'}&=&\Phi(\theta_{\N})',\\
\frac{\partial M_{n3}^{\R}(\theta_{\N},0,0)}{\partial \theta'}&=&\Phi(\theta_{\N}),\\
\frac{\partial M_{n3}^{\R}(\theta_{\N},0,0)}{\partial \lambda'}&=&0,\\
\frac{\partial M_{n3}^{\R}(\theta_{\N},0,0)}{\partial \tau'}&=&0.
\end{eqnarray*}
Expanding $M_{nj}(\hat\theta_{\SEL}^{\R},\hat{\lambda}_{\SEL}^{\R},\hat{\tau}_{\SEL}^{\R})$ at $(\theta_{\N},0,0)$ yields
$$
\left(\begin{array}{ccccc}
-W_2&\Gamma&0\\\Gamma'&0&\Phi'\\
0&\Phi&0\end{array}\right)
\left(\begin{array}{ccccc}\hat{\lambda}_{\SEL}^{\R}\\
\hat\theta_{\SEL}^{\R} - \theta_{\N}\\\hat{\tau}_{\SEL}^{\R}\end{array}\right) = \left(\begin{array}{ccccc}
-M_{n1}^{\R}(\theta_{\N},0,0)\\
0\\0\end{array}\right)+o_p(n^{-1/2}),
$$
where  $\Phi=\Phi(\theta_{\N})$.
Using similar arguments to the proof of Theorem 3, we have
\begin{eqnarray*}
\hat{\theta}_{\SEL}^{\R}-\theta_{\N} &=&  -P_3^{\R}\Gamma'W_2^{-1}M_{n1}^{\R}(\theta_{\N},0,0)+o_p(n^{-1/2}) \,, \\
\hat{\lambda}_{\SEL}^{\R} &=& P_4^{\R}M_{n1}^{\R}(\theta_{\N},0,0)+o_p(n^{-1/2}) \,, \\
\hat{\tau}_{\SEL}^{\R} &=& (\Phi\Sigma_2 \Phi')^{-1}\Phi\Sigma_2\Gamma'W_2^{-1}M_{n1}^{\R}(\theta_{\N},0,0)+o_p(n^{-1/2}) \,,
\end{eqnarray*}
where $P_3^{\R}=\Sigma_2-\Sigma_2 \Phi'(\Phi\Sigma_2 \Phi')^{-1}\Phi\Sigma_2$ and $P_4^{\R} = W_2^{-1}-W_2^{-1}\Gamma P_3^{\R}\Gamma'W_2^{-1}$. It is easy to see that $P_4^{\R}W_2 P_4^{\R}=P_4^{\R}$. Applying the Taylor series expansion, we have that
\[\begin{array}{lllll}
-2nr_{\SEL}(\hat\theta_{\SEL}^{\R},\hat\lambda_{\SEL}^{\R})& =&
2  \sum\limits_{i\in \mS} \log\bigl\{1+\hat{\lambda}_{\SEL}^{\R}w_if_{\N} g_i(\hat\theta_{\SEL}^{\R})\bigr\} \\
 &=& n\hat{\lambda}_{\SEL}^{*\top}W_2\hat{\lambda}_{\SEL}^{\R}+o_p(1)\\
&=&nM_{n1}'(\theta_{\N},0)P_4^{\R}W_2 P_4^{\R}M_{n1}(\theta_{\N},0)+o_p(1)\\
&=&nM_{n1}'(\theta_{\N},0)P_4^{\R}M_{n1}(\theta_{\N},0)+o_p(1).
\end{array}
\]
From the proof of Theorem 5, we have
\[
-2nr_{\SEL}(\hat\theta_{\SEL},\hat\lambda_{\SEL})
=nM_{n1}'(\theta_{\N},0)P_2M_{n1}(\theta_{\N},0)+o_p(1).
\]
It follows that 
\[\begin{array}{lllll}
&& {\rm LR}_{\SEL}^{\R}(\hat\theta_{\SEL}^{\R})  \\
&=& 2n\big\{r_{\SEL}(\hat\theta_{\SEL},\hat\lambda_{\SEL})
-r_{\SEL}(\hat\theta_{\SEL}^{\R},\hat\lambda_{\SEL}^{\R})\big\}\\
&=&nM_{n1}'(\theta_{\N},0)(P_4^{\R}-P_2)M_{n1}(\theta_{\N},0)
+o_p(1) \,\\
&=&nM_{n1}'(\theta_{\N},0)W_2^{-1}\Gamma(\Sigma_2-P_3^{\R})\Gamma'W_2^{-1}M_{n1}(\theta_{\N},0)+o_p(1)\\
&=&nM_{n1}'(\theta_{\N},0)W_2^{-1}\Gamma\Sigma_2 \Phi'(\Phi\Sigma_2 \Phi')^{-1}\Phi\Sigma_2\Gamma'W_2^{-1}M_{n1}(\theta_{\N},0)
+o_p(1).
\end{array}
\]
By Assumption 1, it can be shown that
$$n^{1/2}M_{n1}(\theta_{\N},0)=\frac{n^{1/2}}{N}\sum_{i\in \mS}w_ig_i(\theta_{\N})\stackrel{{\cal L}}{\rightarrow}N(0,\Omega),$$
where $\Omega=(n/N^{2})Var\{\sum_{i\in \mS}w_ig_i(\theta_{\N})\mid {\mF}_{\N}\}$.
Therefore,
\[\begin{array}{lllll}
{\rm LR}_{\SEL}^{\R}(\hat\theta_{\SEL}^{\R}) \stackrel{{\cal L}}{\rightarrow}Q'\Delta_2^{\R}Q,
\end{array}
\]
where
$
\Delta_2^{\R} = \Omega^{1/2}W_2^{-1}\Gamma\Sigma_2 \Phi'(\Phi\Sigma_2 \Phi')^{-1}\Phi\Sigma_2\Gamma'W_2^{-1}\Omega^{1/2}
$
with
$\Sigma_2=(\Gamma' W_2^{-1}\Gamma)^{-1}$. This completes the proof.

\medskip

\no 
{\bf   A.8 \ Theoretical justification of the bootstrap method}

\medskip

\no
The justification of the bootstrap method essentially involves establishing the bootstrap version of Theorem 5.
We consider cases where the final survey weights $w_i$ are calibrated over the known population totals of the $x$ variables using the chi-square distance $D(w,d)$.  The calibrated weights are given by
$w_i=d_i\{1+x_i'\lambda_{c}\}$,
where $\lambda_{c}=(\sum_{i\in \mS}d_ix_ix_i')^{-1}(T_{x}-\hat{T}_{x\HT})$, $T_{x} = \sum_{i=1}^{N}x_i$ and $\hat{T}_{x\HT} = \sum_{i\in \mS}d_ix_i$.
Let $B(\theta)=(\sum_{i=1}^Nx_ix_i')^{-1}\sum_{i=1}^Nx_ig_i(\theta)'$ and $\hat{B}(\theta)=(\sum_{i\in \mS}d_ix_ix_i')^{-1}\sum_{i\in \mS}d_ix_ig_i(\theta)'$.
Under regularity conditions similar to Assumptions 3-5 on the original survey design, we have $\|\hat{B}(\theta)-B(\theta)\|=O_p(n^{-1/2})$ uniformly for all $\theta\in\Theta$. Consequently, we have the following asymptotic expansion:
\begin{eqnarray*}
\hat{U}_n(\theta_{\N})
&=& \sum_{i\in \mS}w_ig_i(\theta_{\N}) \\
&=& \sum_{i\in \mS}d_i[g_i(\theta_{\N})+g_i(\theta_{\N})x_i'\lambda_{c}]\\
&=& \sum_{i\in \mS}d_ig_i(\theta_{\N})+\hat{B}(\theta_{\N})'(T_{x}-\hat{T}_{x\HT})\\
&=& \sum_{i\in \mS}d_ig_i(\theta_{\N})+B(\theta_{\N})'(T_{x}-\hat{T}_{x\HT})+o_p(Nn^{-1/2})\\
&=& \sum_{i\in \mS}d_i[g_i(\theta_{\N})-B(\theta_{\N})'x_i]+B(\theta_{\N})'T_{x}+o_p(Nn^{-1/2}) \,.
\end{eqnarray*}
Let $\hat{U}_n^c(\theta)=\sum_{i\in \mS}d_ie_i(\theta)+B(\theta)'T_{x}$, where $e_i(\theta)=g_i(\theta)-B(\theta)'x_i$. We have
\begin{eqnarray*}
{\rm LR}_{\SEL}(\theta_{\N})
&=&  2\big\{r_{\SEL}(\hat\theta_{\SEL})-r_{\SEL}(\theta_{\N})\big\} \\
&=& nN^{-2}\hat{U}_n^c(\theta_{\N})'W_2^{-1}\Gamma\Sigma_2  \Gamma'W_2^{-1}\hat{U}_n^c(\theta_{\N})+o_p(1) \,.
\end{eqnarray*}

The bootstrap weights $\{w_i^*,\in \mS^*\}$ are created by the same calibration procedure with $T_x$ replaced by $\hat{T}_{x\HT}$. Using similar arguments for the asymptotic expansion to ${\rm LR}_{\SEL}(\theta_{\N})$ and conditional on the original sample, we have a similar expansion to the bootstrap version of the SEL ratio statistic as
\begin{eqnarray*}
{\rm LR}_{\SEL}^*(\hat{\theta}_{\SEL}) &=& 2\big\{r_{\SEL}^*(\hat\theta_{\SEL}^*)-r_{\SEL}^*(\hat\theta_{\SEL})\big\}\\
&=& nN^{-2}\hat{U}_n^{c*}(\hat\theta_{\SEL})'W_2^{-1}\Gamma\Sigma_2  \Gamma'W_2^{-1}\hat{U}_n^{c*}(\hat\theta_{\SEL})+o_p(1) \,,
\end{eqnarray*}
where $\hat{U}_n^{c*}(\theta) = \sum_{i\in \mS^*}d_i^*e_i^*(\theta)+B(\theta)'\hat{T}_{x\HT}$, and $e_i^*(\theta)=g_i^*(\theta)-B(\theta)'x_i^*$ for $i\in \mS^*$. To justify the proposed bootstrap calibration method, it suffices to show that, as $n\rightarrow\infty$,
\[
Var\bigg\{\sum_{i\in \mS}d_ie_i(\theta_{\N})\mid {\mF}_{\N}\bigg\}/Var\bigg\{\sum_{i\in \mS^*}d_i^*e_i^*(\hat\theta_{\SEL})\mid \mS\bigg\}
\longrightarrow 1 \,,
\]
where $Var(\cdot\mid \mF)$ represents the design-based variance and $Var(\cdot\mid \mS)$ denotes the variance under the bootstrap sampling procedure, conditional on the original survey sample $\mS$.

Let  $\hat{\eta}=\sum_{i\in \mS}d_ie_i(\theta_{\N})$,  $\hat{\eta}^*=\sum_{i\in \mS^*}d_i^*e_i^*(\hat\theta_{\SEL})$ and let  $z_i=(nd_i)^{-1}$ and $z_i^*=(nd_i^{*})^{-1}$. We can rewrite $\hat{\eta}$ and $\hat{\eta}^*$ as $\hat{\eta}=n^{-1}\sum_{i\in \mS}\hat{r}_i$ and $\hat{\eta}^*=n^{-1}\sum_{i\in \mS^*}\hat{r}_i^*$, respectively, where $\hat{r}_i^*=e_i^*(\hat\theta_{\SEL})/z_i^*$ and $\hat{r}_i=e_i(\theta_{\N})/z_i$.
Under the proposed with-replacement bootstrap procedure, we have $Var(\hat{\eta}^*\mid \mS)=S_r^2/n$, where $S_r^2=n^{-1}\sum_{i\in \mS}(r_i-\hat{\eta})(r_i-\hat{\eta})'$.
If the original survey sample is selected by single-stage PPS sampling with replacement method, then $\hat{\eta}=n^{-1}\sum_{i\in \mS}g_i(\theta_{\N})/z_i$ is the standard Hansen-Hurwitz estimator and the design-based variance $Var(\hat{\eta}\mid {\mF}_{\N})$ can be unbiasedly estimated by $n^{-1}\{(n-1)^{-1}\}\sum_{i\in \mS}(r_i-\hat{\eta})(r_i-\hat{\eta})'$. It follows that $Var(\hat{\eta}\mid {\mF}_{\N})/Var(\hat{\eta}^*\mid \mS^*) \rightarrow 1$ as $n\rightarrow \infty$. The result also applies to single-stage PPS sampling without replacement with small sampling fractions as commonly used in survey practice on variance estimation.

\medskip

\no 
{\bf   A.9 \ Additional simulation results}

\medskip

\no
Tables \ref{tab1b} and \ref{tab2b} summarize the results on the size and power of the tests for $H_0$: $\theta_{\None} = 1.0$ versus $H_1$: $\theta_{\None} = b$ for PEL and SEL, respectively, with $n/N = 10\%$. The results for $b=1.0$ correspond to the size of the test with nominal value $0.05$ and the results for $b\neq 1.0$ represent the actual power of the test.
Tables \ref{tab3b} and \ref{tab4b} summarize the results on the size and power of the tests for $H_0$: $\theta_{\None}=\theta_{\Ntwo}$ versus $H_1$: $(\theta_{\None},\theta_{\Ntwo})=(b_1,b_2)$, and again with $n/N = 10\%$. The results for $(b_1,b_2)=(1.0,1.0)$ correspond to the size of the test and the results for other values of $(b_1,b_2)$ represent the power of the test.

We further investigate the performance of the empirical likelihood methods for parameters defined through nonsmooth estimating functions. We consider the finite population quantiles and construct empirical likelihood ratio confidence intervals using the proposed methods. The finite population values $\{(x_{1i},x_{2i},y_i), i=1,\cdots,N\}$ for the simulation study are generated from the  superpopulation   model:
$
y_i = 0.5+x_{1i}+x_{2i}+\varepsilon_i,
$
where
$x_{1i} \sim$ Bernoulli$(0.5)$,
$x_{2i} \sim$ Expomential(1) and $\varepsilon_i \sim\chi^2(3)$.
The $100\tau$th finite population quantile $\theta_{\N}(\tau)$  is given by the solution to
$\sum_{i=1}^{N} g_{\tau}(y_i,\theta) = 0$, where $g_{\tau}(y_i,\theta) =
I(y \leq\theta)-\tau$ and $I(\cdot)$  is the indicator function.
We consider scenario A discussed in the previous simulation for creating the final survey weights. We  examine five different quantile
levels at $\tau = 0.10, 0.25, 0.50, 0.75$ and $0.90$.
Three methods are employed to construct   the $95\%$  confidence interval for $\theta_{\N}(\tau)$: the PEL approach, the SEL approach, and the  normal approximation (NA) approach.

 The simulated average length (AL), the coverage probability (CP), the lower tail error (LE) and the upper tail error (UE) rates for the confidence interval $(\hat{\theta}_{\L}\,, \; \hat{\theta}_{\U})$ of parameter $\theta_{\N}(\tau)$ are computed as
\begin{eqnarray*}
&& \mbox{AL} = K^{-1}\sum_{k=1}^K \Bigl\{\hat{\theta}_{\U}^{(k)} - \hat{\theta}_{\L}^{(k)}\Bigr\} \,, \;\;\;\;\;\;\;\;
\mbox{CP} = K^{-1}\sum_{k=1}^K I\Bigl\{\hat{\theta}_{\L}^{(k)} < \theta_{\N}(\tau) < \hat{\theta}_{\U}^{(k)}\Bigr\} \,, \\
&& \mbox{LE} = K^{-1}\sum_{k=1}^K I\Bigl\{\theta_{\N}(\tau) \leq  \hat{\theta}_{\L}^{(k)} \Bigr\} \,, \;\;\;\;
\mbox{UE} = K^{-1}\sum_{k=1}^K I\Bigl\{ \theta_{\N}(\tau) \geq \hat{\theta}_{\U}^{(k)}\Bigr\} \,, \;\;\;\;\;\;\;\;\;\;\;
\end{eqnarray*}
where $(\hat{\theta}_{\L}^{(k)}\,, \; \hat{\theta}_{\U}^{(k)})$ is the confidence interval computed from the $k$th simulation sample,  and $K$ is the total number of simulation runs.

Simulation results based on $B=500$ sets of bootstrap replication weights and $K=1,000$ simulation runs are presented in Table \ref{tab5b}. We have the following major observations: (1) Both the pseudo and the sample empirical likelihood approaches lead to excellent confidence intervals for quantiles in terms of coverage probabilities. (2) The SEL approach gives more   balanced tail error rates than the PEL approach in most cases. (3)  The Wald-type confidence intervals have lower coverages, especially for small or large quantiles (i.e., $\tau = 0.10$ and $0.90$). (4) The pseudo and sample empirical likelihood ratio confidence intervals are slightly wider than the Wald-type intervals for small or large quantiles.

\medskip

\no 
{\bf   A.10 \ Further details of the General Social Survey}

\medskip

\no
The 2016 General Social Survey of Statistics Canada focused on Canadians at Work and Home. The survey questionnaire contained more than 200 questions. The 15 variables used in the application reported in the main paper are derived from the original questions listed below. 

\begin{itemize}

\item {\bf Job Satisfaction (JSR-02)} ~ In general, how satisfied are you with
your job?
\begin{itemize}
\item[]
1: Very satisfied
\item[]
2: Satisfied
\item[]
3: Neither satisfied nor dissatisfied
\item[]
4: Dissatisfied
\item[]
5: Very dissatisfied
\end{itemize}

\item {\bf Sex of Respondent (SEX)} 
\begin{itemize}
\item[]
1: Male
\item[]
2: Female

\end{itemize}

\item {\bf Marital Status of the
Respondent (MARSTAT)} ~ What is your marital status? 
Are you...?
\begin{itemize}
\item[]
1: Married
\item[]
2: Living common-law
\item[]
3: Widowed
\item[]
4: Separated
\item[]
5: Divorced
\item[]
6: Single, never married

\end{itemize}

 \item {\bf Landed Immigrant Status (BPR-16)} ~ Are you now, or have you ever been a
landed immigrant in Canada?
\begin{itemize}
\item[]  1: Yes
\item[]  2: No
 \end{itemize}
 
  \item {\bf Citizenship Status (DCIT)}
\begin{itemize}
\item[]  
1: Canadian citizen by birth only 
\item[] 
2: Canadian citizen by birth and othercitizenship(s) 
\item[] 
3: Canadian citizen by naturalization only 
\item[] 
4: Canadian citizen by naturalization andother
citizenship(s)
\item[] 
5: Other citizenship(s) non-Canadian only 
\item[] 
6: Undetermined

 \end{itemize}
 
\item {\bf  Number of Weeks Employed - Past 12 Months (WET-110) } ~ For how many weeks during the past 12
months were you employed?
\begin{itemize}
\item[] $1,~2,\cdots,~52$
 \end{itemize}

 \item {\bf 
 Number of Weeks Worked at
the Job - Past 12 Months (WLY-145)} ~ During the past 12 months, for how
many weeks did you work at this job?
\begin{itemize}
\item[] $1,~2,\cdots,~52$
 \end{itemize}

 \item {\bf  Unionized Job or Covered
by Contract or Collective
Agreement (WLY-160)}
\begin{itemize}
\item[]  1: Yes
\item[]  2: No
 \end{itemize}

 \item {\bf  Being Happy When Working
Hard (WER-01)} ~ On a scale from 0 to 10, where 0 being
’completely disagree’ and 10 being
’completely agree’, how do you feel
about the following statements? I am
happiest when I work hard.
\begin{itemize}
\item[]  
00: Completely disagree
\item[]  
01 -- 09
\item[]  
10: Completely agree
 \end{itemize}
 
 \item {\bf Employment Benefits -
Workplace Pension Plan (CAB-01A)} ~ Which of the following employment
benefits do you have access to as part of
your employment? - Workplace pension
plan
\begin{itemize}
\item[]  1: Yes
\item[]  2: No
 \end{itemize}
 
  \item {\bf Employment Benefits -
Paid
Sick Leave (CAB-01B)} ~ Which of the following employment
benefits do you have access to as part of
your employment? - Paid
sick leave
\begin{itemize}
\item[]  1: Yes
\item[]  2: No
 \end{itemize}
 
  \item {\bf Employment Benefits -
Paid
Vacation Leave (CAB-01C)} ~ Which of the following employment
benefits do you have access to as part of
your employment? - Paid
vacation leave\begin{itemize}
\item[]  1: Yes
\item[]  2: No
 \end{itemize}
 
 \item {\bf Unfair Treatment/
Discrimination - Past 12
Months (DBH-01)} ~ In the past 12 months, have you
experienced unfair treatment or
discrimination while at work?
\begin{itemize}
\item[]  1: Yes
\item[]  2: No 
 \end{itemize}

\item  {\bf Age Group (AGEARR10)} ~ Age group of the respondent
when came to live
permanently in Canada:
\begin{itemize}
\item[]
1: 15 to 24 years
\item[]
2: 25 to 34 years
\item[]
3: 35 to 44 years
\item[]
4: 45 to 54 years
\item[]
5: 55 to 64 years
\item[]
6: 65 to 74 years
\item[]
7: 75 years and over

\end{itemize}

 \item {\bf  Number of Persons Employed
at Work Location (WLY-147)} ~ About how many persons are employed
at the location where you [work/worked]?
\begin{itemize}
\item[]  
1: Less than 5
\item[]  
2: Between 5 and 19
\item[]  
3: Between 20 and 49
\item[]  
4: Between 50 and 99
\item[]  
5: Between 100 and 500
\item[]  
6: Over 500
 \end{itemize}
 
\end{itemize}

\bigskip

\bigskip

\no {\bf References}

\bigskip

\def\beginref{\begingroup
                \clubpenalty=10000
                \widowpenalty=10000
                \normalbaselines\parindent 0pt
                \parskip.0\baselineskip
                \everypar{\hangindent1em}}
\def\endref{\par\endgroup}

\beginref

Berger, Y. G. and De La Riva Torres, O. (2016). Empirical likelihood confidence intervals for complex sampling designs. \textit{Journal of Royal Statistical Society, Ser. B}, {\bf 78}, 319--341.

Binder, D. A. (1983). The the variances of asymptotically normal estimators from complex surveys. {\em International Statistical Review}, {\bf 51}, 279--292.

Binder, D. A. and Patak, Z. (1994). Use of estimating functions for estimation from complex surveys. {\em Journal of the American Statistical Association}, {\bf 89}, 1035--1043.

Chen, J.  and Qin, J. (1993). Empirical likelihood estimation for finite populations and the effective usage of auxiliary information. {\em Biometrika}, {\bf 80}, 107--116.

Chen, J.  and Sitter, R. R. (1999). A pseudo empirical likelihood approach to the effective use of auxiliary information in complex surveys. {\em Statistica Sinica}, {\bf 9}, 385--406.

Chen, J., Sitter, R. R.  and Wu, C. (2002). Using empirical likelihood methods to obtain range restricted weights in regression estimators for surveys. {\em Biometrika}, {\bf 89}, 230--237.

Chen, S. and Kim, J. K.  (2014). Population empirical likelihood for nonparametric inference in survey sampling. {\em Statistica Sinica}, {\bf  24}, 335--355.

Fan, J. and Li, R. (2001). Variable selection via nonconcave penalized likelihood and its oracle properties. {\em Journal of the American Statistical Association}, {\bf 96}, 1348--1360.

Goodman, R. and Kish, L. (1950). Controlled selection - a technique in probability sampling. {\em Journal of the American Statistical Association}, {\bf  45}, 350--372.

Hartley, H. O. and Rao, J. N. K. (1962). Sampling with unequal probabilities and without replacement. {\em Annals of Mathematical Statistics}, {\bf 33}, 350--374.

Hartley, H. O. and Rao, J. N. K.  (1968). A new estimation theory for sample surveys. {\em Biometrika}, {\bf 55}, 547--557.

Isaki, C. T. and Fuller, W. A. (1982). Survey designs under the regression superpopulation model. {\em Journal of the
American Statistical Association}, {\bf 77}, 89--96.

Kim, J. K. and Wu, C. (2013). Sparse and efficient replication variance estimation for complex surveys. {\em Survey
Methodology}, {\bf 39}, 91--120.

Oguz-Alper, M. and Berger, Y. G. (2016). Modelling complex survey data with population level information: An
empirical likelihood approach. {\em Biometrika}, {\bf 103}, 447--459.

Owen, A. B. (1988). Empirical likelihood ratio confidence intervals for a single functional. {\em Biometrika}, {\bf 75}, 237--249.

Qin, J. and Lawless, J. F. (1994). Empirical likelihood and general estimating equations. {\em The Annals of Statistics}, {\bf 
22}, 300--325.

Rao, J. N. K. and Soctt, A. (1981). The analysis of categorical data from complex sample surveys: chi-squared
tests for goodness-of-fit and independence in two-way tables. {\em Journal of the American Statistical Association}, {\bf 76},
221--230.

Rao, J. N. K. and Scott, A. (1984). On chi-squared tests for multi-way tables with cell proportions estimated from
survey data. {\em Annals of Statistics}, {\bf 12}, 46--60.

Rao, J. N. K.  and  Wu, C. (2009). Empirical likelihood methods. {\em Handbook of Statistics}, Volume 29B, {\em Sample
Surveys: Inference and Analysis}, edited by D. Pfeffermann and C. R. Rao , 189--207.

Rao, J. N. K. and Wu, C. (2010a). Pseudo empirical likelihood inference for multiple frame surveys. {\em Journal of the
American Statistical Association}, {\bf 105}, 1494--1503.

Rao, J. N. K. and Wu, C. (2010b). Bayesian pseudo-empirical-likelihood intervals for complex surveys. {\em Journal of
the Royal Statistical Society, Ser. B},  {\bf 72}, 533--544.

Rao, J. N. K. and Wu, C. F. J. (1988). Resampling inference with complex survey data. {\em Journal of the American
Statistical Association}, {\bf 83}, 231--241.

Rao, J. N. K., Wu, C. F. J. and Yue, K. (1992). Some recent work on resampling methods for complex surveys.
{\em Survey Methodology}, {\bf 18}, 209--217.

Reid, N. (2012). Likelihood inference in complex settings. {\em The Canadian Journal of Statistics}, {\bf 40}, 731--744.

Wang, H., Li, R. and Tsai, C. L. (2007). Tuning parameter selectors for the smoothly clipped absolute deviation method. {\em Biometrika}, {\bf 94}, 553--568.

Wu, C. (2005). Algorithms and r codes for the pseudo empirical likelihood methods in survey sampling. {\em Survey
Methodology}, {\bf 31}, 239--243.

Wu, C. and Lu, W. W. (2016). Calibration weighting methods for complex surveys. {\em International Statistical Review}, 
{\bf 84}, 79--98.

Wu, C. and Rao, J. N. K. (2006). Pseudo-empirical likelihood ratio confidence intervals for complex surveys. {\em The
Canadian Journal of Statistics}, {\bf 34}, 359--375.

Zhao, P. and Wu, C. (2019). Some theoretical and practical aspects of empirical likelihood methods for complex surveys. {\em International Statistical Review}, {\bf 87}, 239--256.

Zhao, P.,  Haziza, D. and Wu, C. (2018). Empirical likelihood inference for complex surveys and the design-based
oracle variable selection theory. Submitted .


Zhong, B. and Rao, J. N. K. (2000). Empirical likelihood inference under stratified random sampling using auxiliary
population information. {\em Biometrika}, {\bf 87}, 929--938.

\endref

\vspace{10cm}

\noindent {\bf End Notes:}

\medskip

\noindent First version: Submitted to {\em Biometrika} on February 19, 2016.

\medskip

\noindent Second version: Submitted to {\em Electronic Journal of Statistics} on March 8, 2019.


\newpage

{\renewcommand{\arraystretch}{1.5}
\tabcolsep0.14 in
\begin{table}[h]
\caption{Size of the PEL and SEL ratio tests assuming standard $\chi^2$ limiting distributions}
\label{tab0}
\begin{tabular}{ccccccccccc}
 &&\multicolumn{3}{c} {PEL} &&\multicolumn{3}{c} {SEL} \\
$n/N$ &  &  $\sigma_1$& $\sigma_2$& $\sigma_3$ && $\sigma_1$& $\sigma_2$& $\sigma_3$\\
& &\multicolumn{9}{c}{$\;\;\;\;\;\;\;\;\;\;\;\;\;\;\;$ $H_0$: $\theta_{\None} = 1.0$ versus $H_1$: $\theta_{\None} = b$}\\
$2\%$ &$b=1.0$ & 0.191 &	0.167 &	0.189 &	& 0.167 &	0.141 &	0.167\\
$10\%$ &$b=1.0$ & 0.194 &	0.170 &	0.193 &	& 0.164 &	0.141 &	0.164\\
& &\multicolumn{9}{c}{$\;\;\;\;\;\;\;\;\;\;\;\;$ $H_0$: $\theta_{\None} = \theta_{\Ntwo}$ versus $H_1$: $(\theta_{\None},\theta_{\Ntwo}) = (b_1,b_2)$}\\
$2\%$  &$b_1=b_2=1.0$ & 0.262 &	0.227 &	0.260 &	& 0.204 &	0.186 &	0.206\\
$10\%$ & $b_1=b_2=1.0$ & 0.261 &	0.248 &	0.264 &	& 0.205 &	0.197 &	0.210\\ 
\end{tabular}
\end{table}
}

{\renewcommand{\arraystretch}{1.3} \tabcolsep0.15 in
\begin{table}
\caption{The Pseudo Empirical Likelihood Approach: Power of the tests for $H_0$: $\theta_{\None} = 1.0$ versus $H_1$: $\theta_{\None} = b$ when $n/N = 2\%$}
\label{tab1}
\begin{tabular}{cccccccccccccccccc}
&& $b=$ & $0.50$ & 0.75& 1.00&1.25& 1.50\\
A&I &$\sigma_1$&	0.991 &	0.647 &	0.046 &	0.637 &	0.994\\
  &&$\sigma_2$&	0.385 &	0.142 &	0.056 &	0.102 &	0.316\\
  &&$\sigma_3$&	0.833 &	0.330 &	0.054 &	0.304 &	0.840\\
  &II &$\sigma_1$&	0.992 &	0.637 &	0.053 &	0.633 &	0.993\\
  &&$\sigma_2$&	0.340 &	0.118 &	0.051 &	0.124 &	0.332\\
  &&$\sigma_3$&0.834 &	0.345 &	0.050 &	0.302 &	0.841\\
  &III &$\sigma_1$&	0.989 &	0.656 &	0.056 &	0.635 &	0.996\\
  &&$\sigma_2$&	0.359 &	0.123 &	0.057 &	0.112 &	0.316\\
  &&$\sigma_3$&	0.836 &	0.320 &	0.054 &	0.300 &	0.840\\
  &IV &$\sigma_1$&	0.986 &	0.654 &	0.050 &	0.638 &	0.996\\
  &&$\sigma_2$&	0.352 &	0.152 &	0.052 &	0.104 &	0.324\\
  &&$\sigma_3$&	0.818 &	0.314 &	0.044 &	0.254 &	0.828\\
  &V &$\sigma_1$&	0.992 &	0.635 &	0.053 &	0.626 &	0.993\\
  &&$\sigma_2$&	0.334 &	0.115 &	0.048 &	0.123 &	0.328\\
  &&$\sigma_3$&	0.831 &	0.343 &	0.049 &	0.300 &	0.834\\
B&I &$\sigma_1$&	0.913 &	0.395 &	0.050 &	0.373 &	0.921\\
  &&$\sigma_2$&	0.226 &	0.094 &	0.040 &	0.087 &	0.181\\
  &&$\sigma_3$&	0.613 &	0.207 &	0.048 &	0.186 &	0.576\\
  &II &$\sigma_1$&	0.916 &	0.403 &	0.055&	0.378 &	0.926\\
  &&$\sigma_2$&	0.199 &	0.079 &	0.038&	0.078 &	0.191\\
  &&$\sigma_3$&0.581 &	0.207 &	0.052&	0.173 &	0.574\\
  &III &$\sigma_1$&	0.923 &	0.413 &	0.049&	0.380 &	0.913\\
  &&$\sigma_2$&	0.202 &	0.099 &	0.053&	0.083 &	0.194\\
  &&$\sigma_3$&0.577 &	0.214 &	0.042&	0.193 &	0.569\\
  &V &$\sigma_1$&	0.921 &	0.416 &	0.059 &	0.397 &	0.929\\
  &&$\sigma_2$&	0.207 &	0.084 &	0.045 &	0.083 &	0.203\\
  &&$\sigma_3$&	0.601 &	0.213 &	0.056 &	0.186 &	0.593\\
\end{tabular}
\end{table}
}

{\renewcommand{\arraystretch}{1.3} \tabcolsep0.15 in
\begin{table}
\caption{The Sample Empirical Likelihood Approach: Power of the tests for $H_0$: $\theta_{\None} = 1.0$ versus $H_1$: $\theta_{\None} = b$ when $n/N = 2\%$}
\label{tab2}
\begin{tabular}{cccccccccccccccccc}
&& $b=$ & $0.50$ & 0.75& 1.00&1.25& 1.50\\
A&I &$\sigma_1$&	0.995 &	0.678 &	0.058 &	0.664 &	0.995\\
  &&$\sigma_2$&	0.393 &	0.150 &	0.059 &	0.116 &	0.331\\
  &&$\sigma_3$&	0.851 &	0.345 &	0.062 &	0.326 &	0.853\\
  &II &$\sigma_1$&	0.994 &	0.674 &	0.062 &	0.665 &	0.994\\
  &&$\sigma_2$&	0.353 &	0.130 &	0.058 &	0.140 &	0.353\\
  &&$\sigma_3$&0.857 &	0.362 &	0.057 &	0.336 &	0.857\\
  &III &$\sigma_1$&	0.995 &	0.667 &	0.059 &	0.664 &	0.996\\
  &&$\sigma_2$&	0.353 &	0.131 &	0.070 &	0.140 &	0.352\\
  &&$\sigma_3$&	0.848 &	0.346 &	0.066 &	0.342 &	0.845\\
  &IV &$\sigma_1$&	0.986 &	0.652 &	0.050 &	0.634 &	0.996\\
  &&$\sigma_2$&	0.344 &	0.150 &	0.046 &	0.100 &	0.316\\
  &&$\sigma_3$&	0.816 &	0.308 &	0.040 &	0.248 &	0.818\\
  &V &$\sigma_1$&	0.992 &	0.635 &	0.053 &	0.626 &	0.993\\
  &&$\sigma_2$&	0.334 &	0.115 &	0.048 &	0.123 &	0.328\\
  &&$\sigma_3$&	0.831 &	0.343 &	0.049 &	0.300 &	0.834\\
B&I &$\sigma_1$&	0.943 &	0.469 &	0.076 &	0.447 &	0.938\\
  &&$\sigma_2$&	0.264 &	0.118 &	0.057 &	0.115 &	0.218\\
  &&$\sigma_3$&	0.662 &	0.245 &	0.066 &	0.236 &	0.630\\
  &II &$\sigma_1$&	0.939 &	0.472 &	0.075&	0.448 &	0.942\\
  &&$\sigma_2$&	0.228 &	0.101 &	0.059&	0.105 &	0.236\\
  &&$\sigma_3$&0.647 &	0.241 &	0.073&	0.225 &	0.632\\
  &III &$\sigma_1$&	0.940 &	0.490 &	0.071&	0.454 &	0.936\\
  &&$\sigma_2$&	0.247 &	0.128 &	0.068&	0.110 &	0.238\\
  &&$\sigma_3$&0.651 &	0.232 &	0.063&	0.227 &	0.631\\
  &V &$\sigma_1$&	0.921 &	0.416 &	0.059 &	0.397 &	0.929\\
  &&$\sigma_2$&	0.207 &	0.084 &	0.045 &	0.083 &	0.203\\
  &&$\sigma_3$&	0.601 &	0.213 &	0.056 &	0.186 &	0.593\\
\end{tabular}
\end{table}
}

{\renewcommand{\arraystretch}{1.3} \tabcolsep0.09 in
\begin{table}
\caption{The Pseudo Empirical Likelihood Approach: Power of the tests for $H_0$: $\theta_{\None} = \theta_{\Ntwo}$ versus $H_1$: $(\theta_{\None},\theta_{\Ntwo}) = (b_1,b_2)$ when $n/N = 2\%$}
\label{tab3}
\begin{tabular}{cccccccccccccccccc}
&& $(b_1,b_2)=$ & $(1.0, 2.0)$ & (1.0, 1.5)& (1.0, 1.0)&(1.5, 1.0)& (2.0, 1.0)\\
A&I &$\sigma_1$&	0.997 &	0.658 &	0.056 &	0.987 &	1.000\\
  &&$\sigma_2$&	0.357 &	0.117 &	0.041 &	0.248 &	0.784\\
  &&$\sigma_3$&	0.866 &	0.306 &	0.055 &	0.754 &	1.000\\
  &II &$\sigma_1$&	0.998 &	0.672 &	0.054 &	0.985 &	1.000\\
  &&$\sigma_2$&	0.339 &	0.129 &	0.053 &	0.263 &	0.779\\
  &&$\sigma_3$&0.876 &	0.312 &	0.055 &	0.762 &	1.000\\
  &III &$\sigma_1$&	0.997 &	0.642 &	0.055 &	0.988 &	1.000\\
  &&$\sigma_2$&	0.339 &	0.113 &	0.062 &	0.255 &	0.779\\
  &&$\sigma_3$&	0.862 &	0.321 &	0.056 &	0.737 &	1.000\\
  &IV &$\sigma_1$&	0.998 &	0.638 &	0.048 &	0.992 &	1.000\\
  &&$\sigma_2$&	0.318 &	0.118 &	0.054 &	0.242 &	0.816\\
  &&$\sigma_3$&	0.822 &	0.272 &	0.040 &	0.742 &	1.000\\
  &V &$\sigma_1$&	0.998 &	0.724 &	0.059 &	0.558 &	0.991\\
  &&$\sigma_2$&	0.364 &	0.148 &	0.050 &	0.110 &	0.320\\
  &&$\sigma_3$&	0.867 &	0.370 &	0.058 &	0.294 &	0.823\\
B&I &$\sigma_1$&	0.931 &	0.381 &	0.051 &	0.864 &	1.000\\
  &&$\sigma_2$&	0.169 &	0.079 &	0.055 &	0.145 &	0.506\\
  &&$\sigma_3$&	0.561 &	0.179 &	0.051 &	0.444 &	0.974\\
  &II &$\sigma_1$&	0.937 &	0.408 &	0.054&	0.863 &	0.999\\
  &&$\sigma_2$&	0.182 &	0.080 &	0.039&	0.143 &	0.465\\
  &&$\sigma_3$&0.599 &	0.177 &	0.058&	0.461 &	0.980\\
  &III &$\sigma_1$&	0.937 &	0.392 &	0.052&	0.855 &	1.000\\
  &&$\sigma_2$&	0.194 &	0.067 &	0.046&	0.134 &	0.516\\
  &&$\sigma_3$&0.593 &	0.177 &	0.053&	0.451 &	0.981\\
  &V &$\sigma_1$&	0.941 &	0.486 &	0.050 &	0.342 &	0.896\\
  &&$\sigma_2$&	0.236 &	0.101 &	0.049 &	0.090 &	0.187\\
  &&$\sigma_3$&	0.638 &	0.232 &	0.053 &	0.174 &	0.570\\
\end{tabular}
\end{table}
}

{\renewcommand{\arraystretch}{1.3} \tabcolsep0.09 in
\begin{table}
\caption{The Sample Empirical Likelihood Approach: Power of the tests for $H_0$: $\theta_{\None} = \theta_{\Ntwo}$ versus $H_1$: $(\theta_{\None},\theta_{\Ntwo}) = (b_1,b_2)$ when $n/N = 2\%$}
\label{tab4}
\begin{tabular}{cccccccccccccccccc}
& & $(b_1,b_2)=$ & $(1.0, 2.0)$ & (1.0, 1.5)& (1.0, 1.0)&(1.5, 1.0)& (2.0, 1.0)\\
A&I &$\sigma_1$&	0.997 &	0.669 &	0.065 &	0.988 &	1.000\\
  &&$\sigma_2$&	0.376 &	0.135 &	0.049 &	0.269 &	0.798\\
  &&$\sigma_3$&	0.865 &	0.322 &	0.064 &	0.771 &	1.000\\
  &II &$\sigma_1$&	0.998 &	0.687 &	0.062 &	0.987 &	1.000\\
  &&$\sigma_2$&	0.360 &	0.140 &	0.061 &	0.288 &	0.797\\
  &&$\sigma_3$&0.878 &	0.332 &	0.063 &	0.778 &	1.000\\
  &III &$\sigma_1$&	0.998 &	0.664 &	0.063 &	0.989 &	1.000\\
  &&$\sigma_2$&	0.361 &	0.131 &	0.069 &	0.278 &	0.789\\
  &&$\sigma_3$&	0.868 &	0.337 &	0.068 &	0.756 &	1.000\\
  &IV &$\sigma_1$&	0.998 &	0.624 &	0.040 &	0.990 &	1.000\\
  &&$\sigma_2$&	0.294 &	0.112 &	0.046 &	0.232 &	0.812\\
  &&$\sigma_3$&	0.798 &	0.252 &	0.038 &	0.732 &	1.000\\
  &V &$\sigma_1$&	0.998 &	0.724 &	0.059 &	0.558 &	0.991\\
  &&$\sigma_2$&	0.364 &	0.148 &	0.050 &	0.110 &	0.320\\
  &&$\sigma_3$&	0.867 &	0.370 &	0.058 &	0.294 &	0.823\\
B&I &$\sigma_1$&	0.941 &	0.445 &	0.076 &	0.897 &	1.000\\
  &&$\sigma_2$&	0.226 &	0.112 &	0.069 &	0.188 &	0.546\\
  &&$\sigma_3$&	0.642 &	0.219 &	0.069 &	0.511 &	0.982\\
  &II &$\sigma_1$&	0.947 &	0.470 &	0.079&	0.897 &	1.000\\
  &&$\sigma_2$&	0.227 &	0.120 &	0.057&	0.192 &	0.522\\
  &&$\sigma_3$&0.645 &	0.219 &	0.079&	0.526 &	0.986\\
  &III &$\sigma_1$&	0.949 &	0.454 &	0.075&	0.893 &	1.000\\
  &&$\sigma_2$&	0.245 &	0.096 &	0.066&	0.178 &	0.577\\
  &&$\sigma_3$&0.644 &	0.234 &	0.077&	0.531 &	0.986\\
  &V &$\sigma_1$&	0.941 &	0.486 &	0.050 &	0.342 &	0.896\\
  &&$\sigma_2$&	0.236 &	0.101 &	0.049 &	0.090 &	0.187\\
  &&$\sigma_3$&	0.638 &	0.232 &	0.053 &	0.174 &	0.570\\
\end{tabular}
\end{table}
}

 {\renewcommand{\arraystretch}{1.5} \tabcolsep0.07 in
\begin{table}[h]
\centering
\caption{GSS Data: Point Estimation, Hypothesis Testing and Variable Selection}
\label{tab-gss}
\begin{tabular}{llcccccccccc}
 \multirow{2}*{Covariate} &\multirow{2}*{Estimate}  & &\multirow{2}*{SE} & &\multirow{2}*{OR} & &\multicolumn{2}{c}{P-Value}  & &\multicolumn{2}{c}{Variable Selection}\\
&  &  &&&  &  & PEL & SEL&& PEL & SEL \\  
1	              &	  -0.029 &&	0.716 &&	0.971  &&	0.967 & 0.962 &&	0.000 &	0.000\\
$x_{1}$	           &	 -0.261 &&	0.211 &&	0.770  &&	0.202 & 0.165 &&	0.000 &	0.000\\
$x_{2}$	           &	 0.211 &&	0.224 &&	1.234  &&	0.342 & 0.312 &&	0.000 &	0.000\\
$x_{3}$	           &	 0.091 &&	0.329 &&	1.095 &&	0.779 & 0.750 &&	0.000 &	0.000\\
$x_{4}$	           &	-0.250 &&	0.258 &&	 0.778  &&	0.319 & 0.305 &&	0.000 &	0.000\\
$x_{5}$	           &	 0.017 &&	0.014 &&	1.017  &&	0.205 & 0.159 &&	0.000 &	0.000\\
$x_{6}$	           &	 -0.012 &&	0.011 &&	0.988  &&	0.274 & 0.261 &&	0.000 &	0.000\\
$x_{7}$	           &	 0.060 &&0.235 &&	1.061  &&	0.792 & 0.785 &&	0.000 &	0.000\\
$x_{8}$	           &	 1.258 &&	0.277 &&	3.518  &&	0.000 & 0.000 &&	2.196 &	2.157\\
$x_{9}$	           &	 0.095 &&	0.268 &&	1.099  &&	0.704 & 0.693 &&	0.000 &	0.000\\
$x_{10}$	           &	 0.590 &&	0.263 &&	1.803  &&	0.019 & 0.013 &&	0.000 &	0.000\\
$x_{11}$	           &	 0.152 &&	0.260 &&	1.164  &&	0.550 & 0.536 &&	0.000 &	0.000\\
$x_{12}$	           &	 -1.422 &&	0.266 &&	0.241  &&	0.000 & 0.676 &&	0.000 &	0.000\\
$x_{13}$	           &	 0.082 &&	0.099 &&	1.085  &&	0.385 & 0.340 &&	0.000 &	0.000\\
$x_{14}$	           &	 0.032 &&	0.059  &&	1.032   &&	0.586 & 0.566 &&	0.000 &	0.000\\
\end{tabular}
\\ 
\vspace{3mm}

Note: The values $0.000$ in the last two columns indicate non-significant factors identified by the variable selection procedure. 
\end{table}
}


{\renewcommand{\arraystretch}{1.3} \tabcolsep0.15 in
\begin{table}
\caption{The Pseudo Empirical Likelihood Approach: Power of the tests for $H_0$: $\theta_{\None} = 1.0$ versus $H_1$: $\theta_{\None} = b$ when $n/N = 10\%$}
\label{tab1b}
\begin{tabular}{cccccccccccccccccc}
&& $b=$ & $0.50$ & 0.75& 1.00&1.25& 1.50\\
A&I &$\sigma_1$&	0.991 &	0.613 &	0.049 &	0.627 &	0.998\\
  &&$\sigma_2$&	0.371 &	0.123 &	0.046 &	0.096 &	0.305\\
  &&$\sigma_3$&	0.822 &	0.312 &	0.046 &	0.302 &	0.850\\
  &II &$\sigma_1$&	0.992 &	0.636 &	0.051 &	0.640 &	0.997\\
  &&$\sigma_2$&	0.361 &	0.141 &	0.046 &	0.102 &	0.314\\
  &&$\sigma_3$&0.843 &	0.331 &	0.043 &	0.307 &	0.846\\
  &III &$\sigma_1$&	0.992 &	0.632 &	0.057 &	0.616 &	0.997\\
  &&$\sigma_2$&	0.358 &	0.141 &	0.057 &	0.108 &	0.312\\
  &&$\sigma_3$&	0.844 &	0.337 &	0.049 &	0.293 &	0.864\\
  &IV &$\sigma_1$&	0.994 &	0.616 &	0.048 &	0.654 &	0.994\\
  &&$\sigma_2$&	0.332 &	0.116 &	0.046 &	0.096 &	0.298\\
  &&$\sigma_3$&	0.802 &	0.306 &	0.048 &	0.288 &	0.808\\
  &V &$\sigma_1$&	0.992 &	0.632 &	0.050 &	0.634 &	0.996\\
  &&$\sigma_2$&	0.358 &	0.139 &	0.044 &	0.102 &	0.305\\
  &&$\sigma_3$&	0.841 &	0.327 &	0.042 &	0.303 &	0.842\\
B&I &$\sigma_1$&	0.911 &	0.435 &	0.043 &	0.365 &	0.917\\
  &&$\sigma_2$&	0.215 &	0.090 &	0.043 &	0.078 &	0.205\\
  &&$\sigma_3$&	0.628 &	0.224 &	0.052 &	0.158 &	0.585\\
  &II &$\sigma_1$&	0.908 &	0.427 &	0.046&	0.380 &	0.924\\
  &&$\sigma_2$&	0.200 &	0.086 &	0.046&	0.064 &	0.196\\
  &&$\sigma_3$&0.604 &	0.209 &	0.049&	0.176 &	0.614\\
  &III &$\sigma_1$&	0.915 &	0.419 &	0.047&	0.359 &	0.915\\
  &&$\sigma_2$&	0.226 &	0.094 &	0.043&	0.078 &	0.183\\
  &&$\sigma_3$&0.613 &	0.204 &	0.045&	0.174 &	0.592\\
  &V &$\sigma_1$&	0.914 &	0.444 &	0.050 &	0.392 &	0.927\\
  &&$\sigma_2$&	0.211 &	0.091 &	0.049 &	0.071 &	0.207\\
  &&$\sigma_3$&	0.628 &	0.220 &	0.057 &	0.191 &	0.635\\
\end{tabular}
\end{table}
}

{\renewcommand{\arraystretch}{1.3} \tabcolsep0.15 in
\begin{table}
\caption{The Sample Empirical Likelihood Approach: Power of the tests for $H_0$: $\theta_{\None} = 1.0$ versus $H_1$: $\theta_{\None} = b$ when $n/N = 10\%$}
\label{tab2b}
\begin{tabular}{cccccccccccccccccc}
&& $b=$ & $0.50$ & 0.75& 1.00&1.25& 1.50\\
A&I &$\sigma_1$&	0.996 &	0.650 &	0.059 &	0.666 &	0.997\\
  &&$\sigma_2$&	0.393 &	0.133 &	0.052 &	0.118 &	0.329\\
  &&$\sigma_3$&	0.855 &	0.332 &	0.057 &	0.341 &	0.864\\
  &II &$\sigma_1$&	0.995 &	0.666 &	0.061 &	0.671 &	0.998\\
  &&$\sigma_2$&	0.374 &	0.151 &	0.056 &	0.118 &	0.345\\
  &&$\sigma_3$&0.871 &	0.346 &	0.052 &	0.353 &	0.856\\
  &III &$\sigma_1$&	0.999 &	0.680 &	0.061 &	0.680 &	0.997\\
  &&$\sigma_2$&	0.372 &	0.144 &	0.054 &	0.124 &	0.347\\
  &&$\sigma_3$&	0.861 &	0.375 &	0.067 &	0.326 &	0.875\\
  &IV &$\sigma_1$&	0.994 &	0.610 &	0.048 &	0.646 &	0.994\\
  &&$\sigma_2$&	0.320 &	0.112 &	0.048 &	0.084 &	0.284\\
  &&$\sigma_3$&	0.798 &	0.304 &	0.044 &	0.286 &	0.806\\
  &V &$\sigma_1$&	0.992 &	0.632 &	0.050 &	0.634 &	0.996\\
  &&$\sigma_2$&	0.358 &	0.139 &	0.044 &	0.102 &	0.305\\
  &&$\sigma_3$&	0.841 &	0.327 &	0.042 &	0.303 &	0.842\\
B&I &$\sigma_1$&	0.947 &	0.507 &	0.071 &	0.437 &	0.943\\
  &&$\sigma_2$&	0.252 &	0.117 &	0.062 &	0.111 &	0.245\\
  &&$\sigma_3$&	0.693 &	0.269 &	0.078 &	0.217 &	0.652\\
  &II &$\sigma_1$&	0.943 &	0.495 &	0.071&	0.466 &	0.943\\
  &&$\sigma_2$&	0.230 &	0.111 &	0.065&	0.096 &	0.248\\
  &&$\sigma_3$&0.676 &	0.257 &	0.076&	0.246 &	0.679\\
  &III &$\sigma_1$&	0.949 &	0.510 &	0.085&	0.442 &	0.941\\
  &&$\sigma_2$&	0.257 &	0.123 &	0.065&	0.108 &	0.228\\
  &&$\sigma_3$&0.682 &	0.270 &	0.081&	0.226 &	0.634\\
  &V &$\sigma_1$&	0.914 &	0.444 &	0.050 &	0.392 &	0.927\\
  &&$\sigma_2$&	0.211 &	0.091 &	0.049 &	0.071 &	0.207\\
  &&$\sigma_3$&	0.628 &	0.220 &	0.057 &	0.191 &	0.635\\
\end{tabular}
\end{table}
}

{\renewcommand{\arraystretch}{1.3} \tabcolsep0.09 in
\begin{table}
\caption{The Pseudo Empirical Likelihood Approach: Power of the tests for $H_0$: $\theta_{\None} = \theta_{\Ntwo}$ versus $H_1$: $(\theta_{\None},\theta_{\Ntwo}) = (b_1,b_2)$ when $n/N = 10\%$}
\label{tab3b}
\begin{tabular}{cccccccccccccccccc}
&& $(b_1,b_2)=$ & $(1.0, 2.0)$ & (1.0, 1.5)& (1.0, 1.0)&(1.5, 1.0)& (2.0, 1.0)\\
A&I &$\sigma_1$&	0.998 &	0.635 &	0.055 &	0.992 &	1.000\\
  &&$\sigma_2$&	0.327 &	0.085 &	0.049 &	0.219 &	0.787\\
  &&$\sigma_3$&	0.880 &	0.301 &	0.048 &	0.748 &	1.000\\
  &II &$\sigma_1$&	0.996 &	0.622 &	0.045 &	0.992 &	1.000\\
  &&$\sigma_2$&	0.306 &	0.099 &	0.041 &	0.229 &	0.761\\
  &&$\sigma_3$&0.897 &	0.314 &	0.044 &	0.758 &	1.000\\
  &III &$\sigma_1$&	0.998 &	0.636 &	0.051 &	0.991 &	1.000\\
  &&$\sigma_2$&	0.306 &	0.110 &	0.046 &	0.239 &	0.780\\
  &&$\sigma_3$&	0.888 &	0.303 &	0.051 &	0.758 &	1.000\\
  &IV &$\sigma_1$&	0.998 &	0.608 &	0.020 &	0.992 &	1.000\\
  &&$\sigma_2$&	0.304 &	0.078 &	0.040 &	0.200 &	0.744\\
  &&$\sigma_3$&	0.906 &	0.308 &	0.048 &	0.714 &	0.998\\
  &V &$\sigma_1$&	0.996 &	0.737 &	0.065 &	0.510 &	0.995\\
  &&$\sigma_2$&	0.294 &	0.094 &	0.047 &	0.132 &	0.358\\
  &&$\sigma_3$&	0.812 &	0.227 &	0.057 &	0.455 &	0.919\\
B&I &$\sigma_1$&	0.924 &	0.366 &	0.051 &	0.842 &	1.000\\
  &&$\sigma_2$&	0.176 &	0.057 &	0.037 &	0.129 &	0.489\\
  &&$\sigma_3$&	0.605 &	0.174 &	0.037 &	0.466 &	0.980\\
  &II &$\sigma_1$&	0.925 &	0.380 &	0.044&	0.862 &	0.999\\
  &&$\sigma_2$&	0.183 &	0.072 &	0.042&	0.145 &	0.473\\
  &&$\sigma_3$&0.607 &	0.176 &	0.046&	0.472 &	0.990\\
  &III &$\sigma_1$&	0.934 &	0.356 &	0.042&	0.857 &	1.000\\
  &&$\sigma_2$&	0.195 &	0.066 &	0.043&	0.149 &	0.477\\
  &&$\sigma_3$&0.585 &	0.161 &	0.053&	0.453 &	0.979\\
  &V &$\sigma_1$&	0.952 &	0.510 &	0.060 &	0.315 &	0.899\\
  &&$\sigma_2$&	0.189 &	0.077 &	0.045 &	0.086 &	0.218\\
  &&$\sigma_3$&	0.545 &	0.155 &	0.055 &	0.298 &	0.717\\
\end{tabular}
\end{table}
}

{\renewcommand{\arraystretch}{1.3} \tabcolsep0.09 in
\begin{table}
\caption{The Sample Empirical Likelihood Approach: Power of the tests for $H_0$: $\theta_{\None} = \theta_{\Ntwo}$ versus $H_1$: $(\theta_{\None},\theta_{\Ntwo}) = (b_1,b_2)$ when $n/N = 10\%$}
\label{tab4b}
\begin{tabular}{cccccccccccccccccc}
&& $(b_1,b_2)=$ & $(1.0, 2.0)$ & (1.0, 1.5)& (1.0, 1.0)&(1.5, 1.0)& (2.0, 1.0)\\
A&I &$\sigma_1$&	0.998 &	0.651 &	0.064 &	0.993 &	1.000\\
  &&$\sigma_2$&	0.355 &	0.108 &	0.057 &	0.252 &	0.804\\
  &&$\sigma_3$&	0.892 &	0.328 &	0.060 &	0.776 &	1.000\\
  &II &$\sigma_1$&	0.996 &	0.635 &	0.055 &	0.991 &	1.000\\
  &&$\sigma_2$&	0.325 &	0.115 &	0.047 &	0.259 &	0.776\\
  &&$\sigma_3$&0.910 &	0.345 &	0.051 &	0.784 &	1.000\\
  &III &$\sigma_1$&	0.997 &	0.644 &	0.061 &	0.994 &	1.000\\
  &&$\sigma_2$&	0.329 &	0.125 &	0.059 &	0.272 &	0.804\\
  &&$\sigma_3$&	0.896 &	0.335 &	0.063 &	0.791 &	1.000\\
  &IV &$\sigma_1$&	0.996 &	0.580 &	0.022 &	0.992 &	1.000\\
  &&$\sigma_2$&	0.282 &	0.080 &	0.034 &	0.184 &	0.734\\
  &&$\sigma_3$&	0.896 &	0.302 &	0.052 &	0.702 &	0.998\\
  &V &$\sigma_1$&	0.996 &	0.737 &	0.065 &	0.510 &	0.995\\
  &&$\sigma_2$&	0.294 &	0.094 &	0.047 &	0.132 &	0.358\\
  &&$\sigma_3$&	0.812 &	0.227 &	0.057 &	0.455 &	0.919\\
B&I &$\sigma_1$&	0.928 &	0.436 &	0.083 &	0.898 &	1.000\\
  &&$\sigma_2$&	0.221 &	0.103 &	0.055 &	0.177 &	0.553\\
  &&$\sigma_3$&	0.668 &	0.218 &	0.077 &	0.545 &	0.991\\
  &II &$\sigma_1$&	0.936 &	0.432 &	0.071&	0.905 &	1.000\\
  &&$\sigma_2$&	0.232 &	0.108 &	0.069&	0.193 &	0.542\\
  &&$\sigma_3$&0.663 &	0.228 &	0.077&	0.556 &	0.995\\
  &III &$\sigma_1$&	0.941 &	0.412 &	0.069&	0.888 &	1.000\\
  &&$\sigma_2$&	0.237 &	0.094 &	0.065&	0.191 &	0.543\\
  &&$\sigma_3$&0.645 &	0.220 &	0.085&	0.527 &	0.987\\
  &V &$\sigma_1$&	0.952 &	0.510 &	0.060 &	0.315 &	0.899\\
  &&$\sigma_2$&	0.189 &	0.077 &	0.045 &	0.086 &	0.218\\
  &&$\sigma_3$&	0.545 &	0.155 &	0.055 &	0.298 &	0.717\\
\end{tabular}
\end{table}
}

{\renewcommand{\arraystretch}{1.4} \tabcolsep0.06 in
\begin{table}[h]
\centering
\caption{The $95\%$ Confidence Intervals for the Finite Population Quantiles}
\label{tab5b}
\begin{tabular}{cccccccccccccccccc}
 &&\multicolumn{4}{c} {$n/N = 2\%$} &&\multicolumn{4}{c} {$n/N = 10\%$}\\
Method&$\tau$&LE&CP&UE&AL&&LE&CP&UE&AL\\ 
PEL&0.10&	0.050& 0.936& 0.014 &0.726 && 0.034 &0.952 &0.014 &0.756 \\
&0.25&	0.024& 0.965 &0.011 &0.646 && 0.029 &0.952 &0.019 &0.661 \\
&0.50&	0.025& 0.959 &0.016 &0.655 && 0.025 &0.954 &0.021 &0.659 \\
&0.75&	0.033 &0.949 &0.018 &0.821 && 0.012 &0.968 &0.020 &0.811 \\
&0.90&	0.028 &0.948 &0.024 &1.109 && 0.019 &0.967 &0.014 &1.176 \\
SEL&0.10&	0.035 &0.944 &0.021 &0.755 && 0.023 &0.957 &0.020 &0.782 \\
&0.25&	0.020 &0.965 &0.015& 0.653 && 0.023 &0.952 &0.025& 0.665 \\
&0.50&	0.020 &0.961 &0.019 &0.656 && 0.021 &0.956 &0.023& 0.659 \\
&0.75&	0.031 &0.950 &0.019 &0.820 && 0.010& 0.967& 0.023& 0.810 \\
&0.90&	0.023 &0.951 &0.026 &1.109 && 0.017& 0.966 &0.017 &1.175 \\
NA&0.10&	0.057 &0.927& 0.016 &0.706 && 0.049 &0.922 &0.029 &0.733 \\
&0.25&	0.035 &0.949 &0.016 &0.643 && 0.037 &0.932& 0.031& 0.645 \\
&0.50&	0.026 &0.941 &0.033 &0.650 && 0.026 &0.939 &0.035 &0.657\\
&0.75&	0.035& 0.934 &0.031 &0.815 && 0.017 &0.954 &0.029& 0.790 \\
&0.90&	0.039 &0.922 &0.039& 1.095 && 0.021 &0.935 &0.044 &1.163\\

\end{tabular}
\end{table}
}

\end{document}